\documentclass[usenatbib]{mnras}

\usepackage{newtxtext,newtxmath}

\usepackage[T1]{fontenc}
\usepackage{amsmath}
\usepackage{scalerel}
\DeclareRobustCommand{\VAN}[3]{#2}
\let\VANthebibliography\thebibliography
\def\thebibliography{\DeclareRobustCommand{\VAN}[3]{##3}\VANthebibliography}

\usepackage{ae,aecompl}
\usepackage{xparse,xspace}
\usepackage{cancel}
\usepackage{siunitx}
\usepackage{graphicx}	
\usepackage{amsmath}	
\usepackage{amsfonts}
\usepackage{graphicx}
\usepackage{color}
\usepackage{bm}
\usepackage{lipsum}
\usepackage{nccmath}
\usepackage{tabularx}
\usepackage{caption}
\usepackage{multicol}
\usepackage{graphicx}
\usepackage[export]{adjustbox}
\usepackage[flushleft]{threeparttable}
\usepackage{tabularray}

\title[Detecting PNS rotation and spindown using SN neutrinos]{Prospects for detecting proto-neutron star rotation and spindown using supernova neutrinos}

\author[Prasanna et al.]{
Tejas Prasanna$^{1,2}$\thanks{E-mail: prasanna.9@osu.edu},
Todd A. Thompson$^{2,3,1}$\thanks{E-mail: thompson.1847@osu.edu}
\& Christopher Hirata $^{1,2,3}$\thanks{E-mail: hirata.10@osu.edu}
\\
$^{1}$Department of Physics, The Ohio State University, Columbus, Ohio 43210, USA\\
$^{2}$Center for Cosmology \& Astro-Particle Physics, The Ohio State University, Columbus, Ohio 43210, USA\\
$^{3}$Department of Astronomy, The Ohio State University, Columbus, Ohio 43210, USA\\}

\date{Accepted XXX. Received YYY; in original form ZZZ}

\pubyear{2023}

\begin{document}

\label{firstpage}
\pagerange{\pageref{firstpage}--\pageref{lastpage}}
\maketitle

\begin{abstract}
After a successful supernova, a proto-neutron star (PNS) cools by emitting neutrinos on $\sim 1-100$\,s timescales. Provided that there are neutrino emission `hot-spots' or `cold-spots' on the surface of the rotating PNS, we can expect a periodic modulation in the number of neutrinos observable by detectors. We show that Fourier transform techniques can be used to determine the PNS rotation rate from the neutrino arrival times. Provided there is no spindown, a 1-parameter Discrete Fourier Transform (DFT) is sufficient to determine the spin period of the PNS. If the PNS is born as a magnetar with polar magnetic field strength $B_0 \gtrsim 10^{15}$\,G and is `slowly' rotating with an initial spin period $\gtrsim 100$\,ms, then it can spindown to periods of the order of seconds during the cooling phase. We propose a modified DFT technique with three frequency parameters to detect spindown. Due to lack of neutrino data from a nearby supernova except the $\sim20$ neutrinos detected from SN1987A, we use toy models and one physically motivated modulating function to generate neutrino arrival times. We use the false alarm rate (FAR) to quantify the significance of the Fourier power spectrum peaks. We show that PNS rotation and spindown are detected with $\rm FAR<2\%$ ($2\sigma$) for periodic signal content $\rm M\gtrsim 13-15\%$ if $5\times 10^{3}$ neutrinos are detected in $\sim 3$\,s and with $\rm FAR<1\%$ for $\rm M\geq 5\%$ if $5\times 10^{4}$ neutrinos are detected in $\sim 3$\,s. Since we can expect $\sim 10^{4}-10^{5}$ neutrino detections from a supernova at 10\,kpc, detection of PNS rotation and spindown is possible using the neutrinos from the next Galactic supernova. 
\end{abstract}
\begin{keywords}
supernovae: general -- stars: neutron -- stars: magnetars -- neutrinos 
\end{keywords}



\section{Introduction}
\label{section:introduction}
When stars with masses $\gtrsim8-10$\,M$_\odot$ explode successfully in a core-collapse supernova, the gravitational binding energy of the remaining neutron star is emitted in neutrinos \citep{Burrows1986}, which drive a thermal wind \citep{Duncan1986, Burrows1995, Janka1996}. The neutrino-driven wind cooling phase lasts $\sim1-100$\,s after the supernova (SN) with the neutrino luminosity decreasing as a function of time (e.g., \citealt{Pons1999, Li2021}). 

The neutrinos carry critical information about the proto-neutron star (PNS) left behind by the explosion. The neutrinos detected from SN1987A have been used to constrain the physics of the explosion \citep{Burrows1987,OConnor2013}, the mass of the PNS \citep{Fiorillo2023}, and the possibility of black hole formation \citep{Kfir2016}. The neutrinos have also been used to constrain microphysics, including the neutrino masses \citep{Arnett1987,Bahcall1987,Burrows1987}, dense-matter phase transitions \citep{Takahara1988,Keil1995}, the existence of axions \citep{Turner1988}, and the possibility of neutrino-neutrino interactions \citep{Chang2023}.

Here, we focus on the possibility that the neutrino signal from the next Galactic supernova might encode and evidence PNS rotation and spindown. Previous works have explored the possibility of identifying the progenitor core rotation using gravitational waves and neutrinos \citep{Yokozawa2015}. From the SN1987A event, $\sim 20$ neutrinos were detected \citep{Hirata1987, Bionta1987}. Several papers suggested a periodicity in neutrino arrival times from SN1987A (e.g., \citealt{Harwit1987}) and some others argued that the periodicity was not statistically significant (e.g., \citealt{Schaefer1988}). With only about 20 neutrino detections, it is difficult to extract PNS rotation rates from the neutrino signal. However, modern neutrino detectors like the Super-Kamiokande can detect about $10^{4}$ neutrinos for a supernova at 10\,kpc (e.g., \citealt{Beacom2000, Beacom2002}). If Betelgeuse explodes as a successful supernova at a distance of $\sim0.2$\,kpc, we can expect about $\sim10^{7}$ neutrino detections at Super-Kamiokande. A larger number of detections can be expected from a bigger detector like the Hyper-Kamiokande. For a supernova at 10\,kpc, Hyper-Kamiokande is expected to detect $\sim 5-9\times 10^{4}$ neutrinos during the cooling phase and for a nearby supernova at the distance of Betelgeuse, the peak event rate is expected to reach $10^{8}$\,Hz \citep{Abe2021}. If periodicity exists in the neutrino signal, then these many neutrinos may be sufficient to extract information about the PNS rotation rate.

The kind of long-lived angular asymmetry in the neutrino emission from PNSs needed for a possible detection of rotation is highly uncertain, but there is no doubt that such analyses would be immediately performed in the event of a Galactic SN, which would be a landmark event in astrophysics and astronomy. Asymmetric neutrino emission is invoked to explain neutron star kicks at birth (e.g., \citealt{Coleman2022}). While uncertain, we are motivated to consider long-lived asymmetries by several pieces of physics. For example, some studies of supernovae (SNe) find that neutrino emission from the PNS is not isotropic. \cite{Tamborra2014} show that asymmetries such as Lepton-number Emission Self-sustained Asymmetry (LESA) arise where the $\rm \nu_e$ and $\bar{\nu}_{\rm e}$ fluxes have a dipole pattern (see also \citealt{Vartanyan2019,Nagakura2021}). Although SN studies have not simulated the post-explosion dynamics throughout the cooling time lasting tens of seconds, it is possible that neutrino emission asymmetries are sustained. Some studies have explored the possibility of detection of the modulation in neutrino events rates (e.g., \citealt{Takiwaki2018}). 

Analogous to starspots, which are localized regions of high magnetic field that can cause quasi periodic brightness changes in rotating stars (see for example \citealt{Basri2020}), it is possible that small scale magnetic fields on the PNS can result in asymmetric neutrino emission. The temperature and density respectively at the neutrinosphere of the PNS are $T_0 \sim 5$\,MeV and $\rho_0 \sim 10^{12}$\,g cm$^{-3}$, leading to a pressure of $\sim 10^{31}$\,ergs cm$^{-3}$ \citep{Thompson2001}. This suggests that small scale magnetic fields of $\sim 10^{16}$\,G can lead to neutrino emission asymmetries that may be analogous to the photospheres of convective stars that exhibit rotational modulation (``rotational variables;'' e.g., \citealt{Phillips2023,Jayasinghe2018}). In addition, \citealt{Arras1999} suggest that large scale surface magnetic fields of $\sim 10^{15}-10^{16}$\,G can result in asymmetric neutrino emission. Provided that there are approximately fixed neutrino emission `hot-spots' or `cold-spots' on the surface of the PNS, we can expect a periodic modulation in neutrino counts per time interval. In this paper, we study how well the PNS rotation rates can be measured from the neutrino arrival times if there is a periodic modulation for various magnitudes of assumed modulation amplitude.

A second reason strong magnetic fields may affect any potential rotational modulation in the neutrino signal from PNSs is if the PNS is endowed with a magnetar-strength magnetic field at birth. Following earlier works by \cite{Thompson2004}, \cite{Bucciantini2006}, and \cite{Metzger2007}, \cite{Prasanna2022} have shown that the PNS might spin down significantly during the cooling epoch. Magnetar birth is common in the Galaxy, representing $\sim 40$\% of neutron star births \citep{Beniamini2019}. There is thus a possibility that the next Galactic supernova may result in a magnetar that spins down during the cooling phase due to strong magnetic fields and angular momentum carried away by the magneto-centrifugal wind. \cite{Prasanna2022} use two-dimensional magnetohydrodynamic simulations to show that a PNS born `slowly' rotating with initial spin period $\gtrsim 100$\,ms and polar magnetic field strength $B_0\gtrsim 10^{15}$\,G can spin down significantly to spin periods of the order of seconds during the cooling phase lasting just a few tens of seconds. They show that slower initial spin periods and larger polar magnetic field strengths lead to faster spindown. They also provide an estimate of the critical value of $B_0$ required for fast spindown for a given initial spin period. For the rare class of rapidly rotating millisecond magnetars with spin period $\lesssim 5$\,ms that might plausibly power gamma-ray bursts (GRBs; \citealt{Metzger2011}) and super luminous supernovae (SLSNe; \citealt{Kasen2010}), spindown is not significant during the cooling epoch \citep{Prasanna2023,Raives2023}.  

The combination of high expected neutrino count rates from the next Galactic SN with the possibility of long-lived neutrino asymmetries and spindown motivate a general exploration of the prospects for detecting the PNS spin period and its potential evolution using neutrinos. In this paper, we show that the rotation rate and spindown of the PNS can be determined using Fourier techniques from the supernova neutrino arrival times provided that there is a periodic modulation in the neutrino signal. As mentioned in the previous paragraph, PNS spin period can evolve significantly during the cooling phase or remain roughly constant as a function of time depending on the spin period of the PNS at birth and the strength of the magnetic field. We show that the standard 1-parameter Discrete Fourier Transform (DFT) is sufficient to detect the spin period of the PNS if the PNS spin period remains constant. However, if the PNS spins down significantly, then the standard DFT is insufficient. We propose a modification to the DFT technique and show that PNS spindown can be detected. In contrast to the one frequency parameter in standard Fourier transforms, our modification has three frequency parameters. 

In Section \ref{method}, we describe the 1-parameter and 3-parameter DFT. We also provide an intuitive explanation of the techniques. In Section \ref{results}, we describe the Fourier power spectrum results and quantify the significance of the power spectrum peaks. We show that the rotation and spindown can be detected with a false alarm rate $\rm (FAR)<2\%$ ($2\sigma$) for periodic signal content $\rm M\gtrsim 13-15\%$ if $5\times 10^{3}$ neutrinos can be detected in $\sim 3$\,s. If $5\times 10^{4}$ neutrinos can be detected in $\sim 3$\,s, we show that PNS rotation and spindown can be detected with $\rm FAR<1\%$ for $\rm M\geq 5\%$.
In Section \ref{conclusions}, we summarise the techniques and the main results of the paper.

\section{Method}
\label{method}

\subsection{Models for generating the neutrino arrival times}
\label{pdfs}
\begin{figure}
\centering{}
\includegraphics[width=\linewidth]{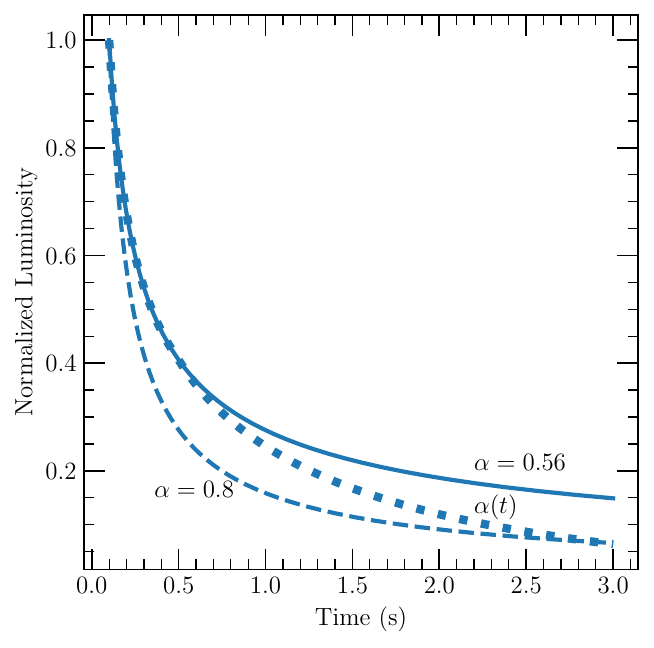}
\caption{Normalized neutrino luminosity as a function of time for three different neutrino luminosity decay coefficients. The solid line and the dashed line show the luminosity profiles for fixed $\alpha=0.56$ and $\alpha=0.8$ respectively. The dotted line shows the luminosity profile for a decay coefficient that is a function of time. We have $\alpha(t)=0.56$ for $0.1$\,s$\le t\le0.5$\,s and $\alpha(t)=0.56+0.1(t-0.5\,{\rm s})$ for $0.5$\,s$<t\le3$\,s.} 
\label{pons_lum}
\end{figure}

Our analysis uses neutrino arrival times at the detectors following a supernova. Since only about 20 neutrinos have been detected from SN1987A \citep{Hirata1987,Bionta1987} and there is no neutrino data from a nearby SN, we use a toy model to generate neutrino arrival times based on a probability distribution. We use the following probability density function (PDF) which gives the probability density that a given neutrino arrived at a time $t$ in the domain $t_i\le t \le t_f$:
\begin{equation}
\label{PDF}
    P(t)=\left[A+BZ\left(\phi_i+\int_{t_i}^t \Omega(t') dt'\right)\right]\left(\frac{t_i}{t}\right)^{\alpha},
\end{equation}
where $Z$ is the periodic modulating function, $t_i>0$ and $t_f$ are the times of arrival of the first and the last neutrino used for the analysis respectively, $A$ and $B$ are constants which are normalized so that $\int_{t_i}^{t_f}P(t)dt=1$, $\phi_i$ is the initial phase of the neutrino signal, $\Omega(t)$ is the PNS angular frequency of rotation as a function of time, and $\alpha\geq0$ is the neutrino luminosity decay coefficient. The domain of $P(t)$, which is $t_i\le t \le t_f$, corresponds to some time interval during the PNS cooling phase when periodic modulation in the neutrino signal exists. $P(t)$ contains a non-periodic part quantified by the coefficient $A$ which is added to a periodic modulation quantified by the the coefficient $B$. The neutrino luminosity decreases as a function of time during the cooling phase (e.g., \citealt{Pons1999, Vartanyan2023}), which is mimicked by the power law decay as a function of time of both the periodic and non-periodic parts in equation \ref{PDF}. Motivated by the \cite{Pons1999} cooling models, we choose three different values of $\alpha$ (of which two are constant in time and one is time-dependent) corresponding to three different neutrino luminosity decay profiles (normalized to unity) as shown in Figure \ref{pons_lum} to test our detection algorithm. Due to the high timing accuracy of modern neutrino detectors, we do not consider the errors in neutrino arrival time measurements.  

We define the periodic signal content M which is directly related to the second term in equation \ref{PDF} as follows:
\begin{equation}
\label{per_cont}
    {\rm M}=B\int_{t_i}^{t_f}Z\left(\phi_i+\int_{t_i}^t \Omega(t') dt'\right)\left(\frac{t_i}{t}\right)^{\alpha}dt.
\end{equation}
M quantifies the fraction of neutrinos contributed by the periodic part. In other words, M is the fraction of neutrinos emitted by the hot-spots on the PNS. 

In this paper, we focus on the case of neutrino emission hot-spots only. We emphasize that the detection algorithm presented in this paper applies equally to the case of the neutrino emission cold-spots. The interpretation of hot-spots is that they enhance the number of neutrinos compared to the background neutrino emission by the PNS when they are visible. In contrast, the cold-spots reduce the number of neutrinos compared to the background when they are visible. Since the detection algorithm presented in this paper detects only the periodicity in the neutrino signal and is not sensitive to whether the neutrino signal is enhanced or diminished, we do not specifically present results for the cold-spots. We note that the coefficient $B$ is positive in the case of hot-spots and negative in the case of cold-spots.

\begin{figure*}
\centering{}
\includegraphics[width=\textwidth]{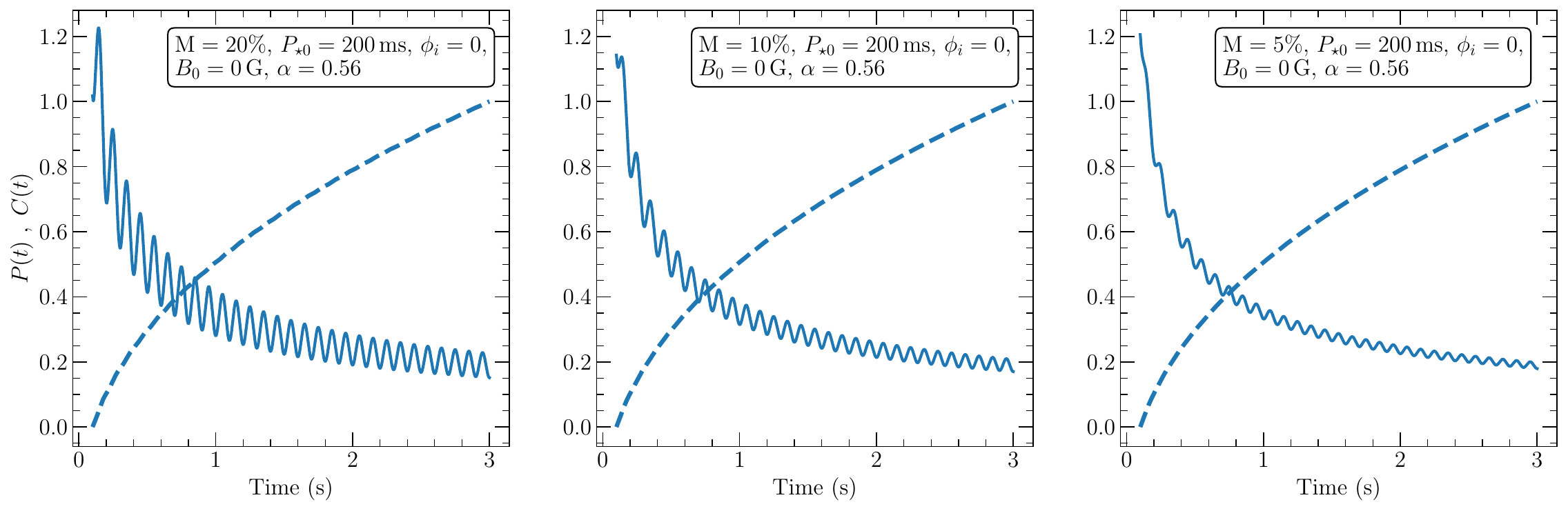}
\caption{PDF $P(t)$ (solid lines) and CDF $C(t)$ (dashed lines) for a constant PNS spin period of 200\,ms with initial phase $\phi_i=0$ for different values of periodic content M. The modulating function is a $\sin^2$ function. The neutrino luminosity decay coefficient $\alpha=0.56$. This modulating function translates to two identical neutrino emission hot-spots on opposite sides of the PNS. We use the label polar magnetic field strength $B_0=0$\,G to imply a constant PNS spin period, because spindown is negligible when the magnetic field strength is $\lesssim 10^{14}$\,G \citep{Prasanna2022, Prasanna2023}.} 
\label{pdf_t3s}
\end{figure*}

The cumulative distribution function (CDF) is defined as
\begin{equation}
\label{CDF}
    C(t)=\int_{t_i}^t P(t')dt'.
\end{equation}
For $P(t)>0$, $C(t)$ is a strictly increasing function between 0 and 1. As an illustration, Figure \ref{pdf_t3s} shows the PDF $P(t)$ (solid lines) and the CDF $C(t)$ (dashed lines) as a function of time for different values of the periodic content M for a PNS with a constant spin period of 200\,ms. In Figure \ref{pdf_t3s}, the modulating function is a $\sin^2$ function, that is, $Z\left(\phi_i+\int_{t_i}^t \Omega(t') dt'\right)=\sin^2\left(\phi_i+\int_{t_i}^t \Omega(t') dt'\right)$. 

\subsubsection{Alternative modulation functions}
\label{alt_mod}
\begin{figure*}
\centering{}
\includegraphics[width=\textwidth]{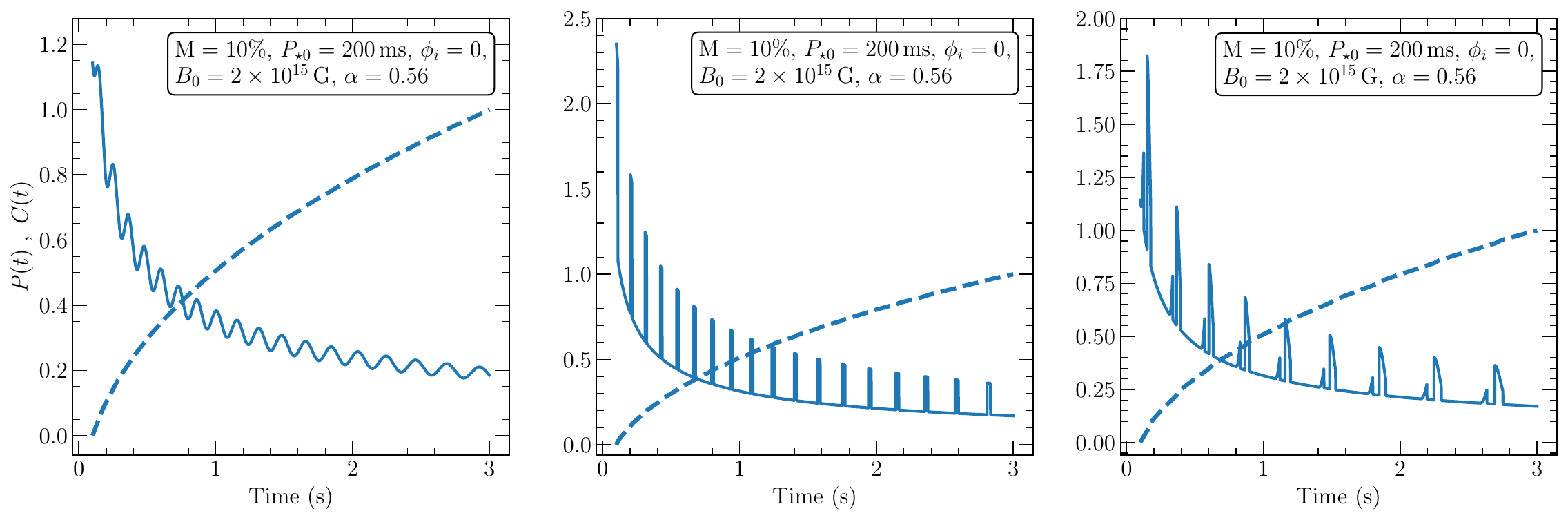}
\caption{PDF $P(t)$ (solid lines) and CDF $C(t)$ (dashed lines) for initial PNS spin period $P_{\star0}=200$\,ms and polar magnetic field strength $B_0=2\times 10^{15}$\,G with initial phase $\phi_i=0$ for periodic content $\rm M=10\%$. The modulating function is a $\sin^2$ function in the left panel, a square wave with a duty cycle of 10\% in the middle panel, and a combination of sinusoidal and square waves in the right panel. Modulating functions in the left and the middle panels translate to two identical neutrino emission hot-spots on opposite sides of the PNS. The modulating function in the right panel corresponds to two different kinds of hot-spots on the PNS. We label the spindown rate with the polar magnetic field strength $B_0$ and the initial spin period $P_{\star0}$. These are based on spindown models from \citealt{Prasanna2022}. However, it is possible that similar spindown rates are achieved with lower values of $B_0$ at later times during the cooling phase that have not been explored in \citealt{Prasanna2022}.} 
\label{pdf_t3s_B2e15}
\end{figure*}

We have run simulations with various modulating functions to show that our detection algorithm works for various potential configurations of neutrino emission hot-spots on the PNS. Table \ref{mod_func_table} lists all the modulating functions used in this paper. Figure \ref{pdf_t3s_B2e15} shows profiles of $P(t)$ and $C(t)$ obtained using three different modulating functions for $\rm M=10\%$ for a PNS with polar magnetic field strength $B_0=2\times 10^{15}$\,G and an initial spin period $P_{\star0}=200$\,ms which spins down to a period of 500\,ms at the end of 2.9\,s of evolution (this is an actual spindown model from \citealt{Prasanna2022}). The start and end times in Figure \ref{pdf_t3s_B2e15} correspond to $t_i=0.1$\,s and $t_f=3$\,s. We choose these values for $t_i$ and $t_f$ for most of the results presented in this paper (refer to Section \ref{const_spin} for the reasons for these choices). In Figure \ref{pdf_t3s_B2e15}, the left panel shows a $\sin^2$ modulating function, the middle panel shows a square wave modulation with 10\% duty cycle, and the right panel shows a modulating function obtained from a combination of sinusoidal and square waves. We note that the square wave modulation with only a 10\% duty cycle may not be physically feasible, but we consider this function to show that detection is possible with extreme mathematical models. The increase in spin period with time is evident in all the panels. The nature of the modulating functions in the left and the middle panels translate to two identical neutrino emission hot-spots on opposite sides of the PNS while the modulating function in the right panel translates to two different kinds of hot-spots. We label the PNS spindown rate with the polar magnetic field strength $B_0$ and the initial spin period $P_{\star0}$ based on the spindown models from \cite{Prasanna2022}. However, it may be possible that similar spindown rates are achieved with lower values of $B_0$ at later times during the cooling phase that have not been explored in \cite{Prasanna2022}. 

\begin{figure}
\centering{}
\includegraphics[width=\linewidth]{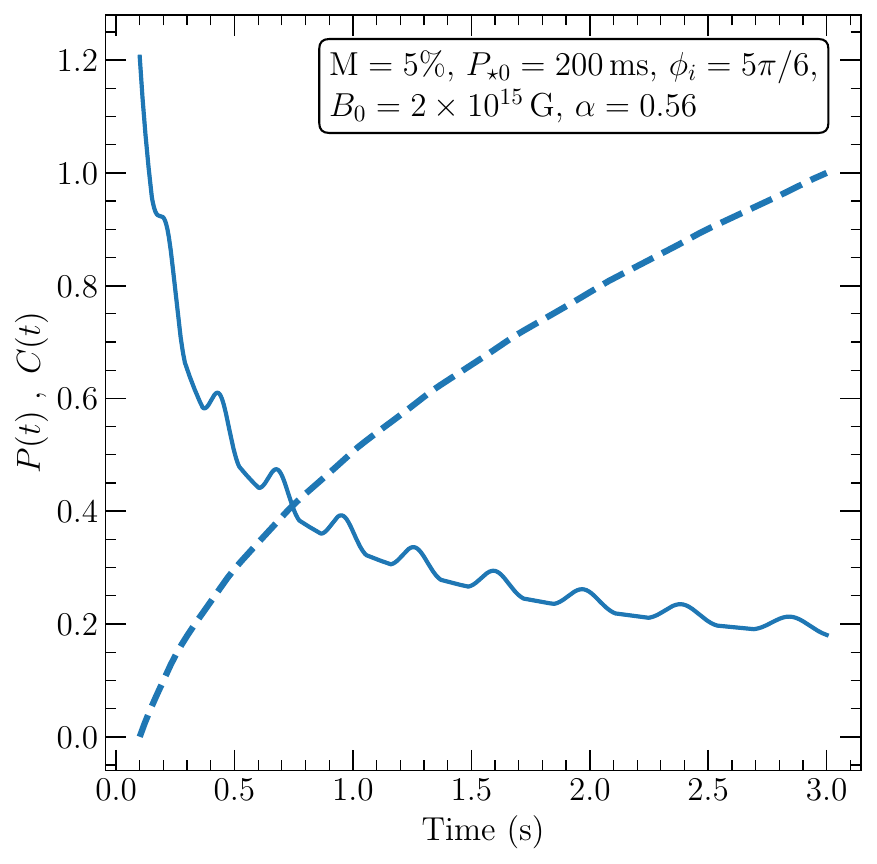}
\caption{PDF $P(t)$ (solid lines) and CDF $C(t)$ (dashed lines) for periodic content $\rm M=5\%$ for the spindown model with initial spin period $P_{\star0}=200$\,ms and polar magnetic field strength $B_0=2\times 10^{15}$\,G (refer to the caption of Figure \ref{pdf_t3s_B2e15} for the meaning of spindown rates labelled in terms of $B_0$ and $P_{\star0}$) for a physically motivated modulation function obtained assuming a small (size $\ll$ PNS neutrinosphere radius) neutrino emission hot-spot on the PNS equator (see Section \ref{alt_mod} for details).} 
\label{pdf_t3s_Chrismod}
\end{figure}

Figure \ref{pdf_t3s_Chrismod} shows the PDF and CDF for $\rm M=5\%$ for the spindown model with $P_{\star0}=200$\,ms and $B_0=2\times 10^{15}$\,G for a physically motivated modulation function with a hot-spot. The modulation has been obtained assuming a small (size $\ll$ PNS neutrinosphere radius $R$) neutrino emission hot-spot at the PNS equator and an observer positioned along the equator for a PNS with a mass of $1.4$\,$\rm M_{\odot}$ and a radius of 12\,km. We model the visibility of the hot-spot using general relativity (GR) because in GR a PNS can ``self-lens'' and hence a hot-spot may be visible even if on the side facing away from the observer, thus leading to a wider modulation function than the Newtonian calculation. We assume that the hot-spot is on a spherical PNS that is much denser than its surroundings and that the external spacetime follows the Schwarzschild geometry, and rotational effects such as frame dragging and Doppler boosting can be neglected. If a neutrino is radiated from the neutrinosphere (radius $R$) at angle $\theta = \cos^{-1}\mu$ to the radial direction, then when it escapes to $\infty$ it does so in the direction:
\begin{equation}
\psi = {\cal L} \int_0^{GM_\star/c^2R} \frac{du}{\sqrt{1 - {\cal L}^2 u^2(1-2u)}},
\end{equation}
where $u=GM_\star/c^2r$ is the normalized inverse radial coordinate (with $u=GM_\star/c^2R$ at the neutrinosphere and $u=0$ when the neutrino has escaped to $\infty$)\footnote{Aside from the normalization by $M_\star$, this follows the definition of $u$ in the treatment of \citet[\S20]{mtbh}. We evaluate the integrals numerically even though they could also be written as incomplete elliptic integrals \citep{1931JJAG....8...67H}.} and the dimensionless neutrino angular momentum is
\begin{equation}
{\cal L} = \frac{c^2R}{GM_\star}\,\sqrt{\frac{1-\mu^2}{1-2GM_\star/c^2R}}.
\end{equation}
Here $\psi=0$ corresponds to a neutrino escaping straight up; $\psi>\pi/2$ may occur due to self-lensing, and $\psi=\pi$ corresponds to a neutrino that escapes in the antipodal direction (which does not happen for our parameters). 

If the hot-spot has uniform surface brightness (no ``neutrino limb darkening''), then the probability distribution for the emission angle is $d{\rm Prob}/d\mu \propto \mu$ (the single power of $\mu=\cos\theta$ is the familiar projection effect: the area element $dA_\perp$ perpendicular to the neutrino ray is related to the surface area element of the neutrinosphere by $dA_\perp = \cos\theta\,dA$). In most real problems, the emerging radiation has the highest surface brightness in the $\mu=1$ direction, and this effect will enhance rotational modulation. We have therefore applied the surface brightness distribution for isotropic scattering, using the 8th order of approximation according to \citet[Eqs.~III.60,61]{Chandrasekhar1960}. Neutrino scattering is actually a mixture of processes with different angular distributions, but we note that the limb darkening law depends only weakly on this detail. 
When visible, the hot-spot subtends a solid angle $d\Omega \propto \sin\psi d\psi$ on the observer sky. We obtain the observed intensity of the hot-spot as a function of rotation phase of the PNS by using the flux per unit solid angle.

The geometry that would lead to the ``sharpest'' modulation function would be for a small hot-spot (size $\ll R$); otherwise there is further finite-size smoothing. Furthermore, the best geometry for detecting modulation is a hot-spot on the PNS equator, and the observer in the PNS equatorial plane (modulation decreases with latitude and vanishes at the poles). We take this optimal configuration in what follows, with the expectation that it represents the sharpest physically reasonable modulating function.
\begin{table*}

	\caption{List of modulating functions used in this paper.}
\label{mod_func_table}
\begin{tabular}[width=\textwidth]{@{}ccc@{}} 
		\hline
		 Modulating function & Nature of hot-spots on the PNS & Reference \\ 
\hline
Squared sinusoid & Two symmetric hot-spots on opposite sides of the PNS & Figure \ref{pdf_t3s} and left panel in Figure \ref{pdf_t3s_B2e15} \\
Two hot-spot square wave & Two symmetric hot-spots on opposite sides of the PNS & middle panel in Figure \ref{pdf_t3s_B2e15} \\
Combination of sinusoid and square waves & Two different kinds of hot-spots placed asymmetrically on the PNS & Right panel in Figure \ref{pdf_t3s_B2e15} \\
Physically motivated modulation & Small (size $\ll$ PNS neutrinosphere radius) hot-spot on the PNS equator & Figure \ref{pdf_t3s_Chrismod} \\
\hline 
	   \end{tabular}
 \end{table*}

\subsubsection{Generating neutrino arrival times from probability distributions}
\label{arrival_times_subsub}
Random numbers following the PDF $P(t)$ can be generated from a set of random numbers drawn from a uniform distribution $U(0,1)$ by just inverting the CDF $C(t)$ \citep{Numerical_recipes}. We compute $P(t)$ at $\rm N_{\rm pdf}$ uniformly spaced points between $t_i$ and $t_f$. We compute $C(t)$ using the $\rm N_{\rm pdf}$ samples of $P(t)$. We then generate a set of $\rm N_{\nu}$ random numbers that follow a uniform distribution $U(0,1)$, where $\rm N_{\nu}$ is the total number of neutrinos. We use the numerical inverse of $C(t)$ to generate the neutrino arrival times from the uniform random numbers. That is, if $0\le x_0<1$ is the random number generated from $U(0,1)$, then the corresponding neutrino arrival time is given by $C^{-1}(x_0)$. If $C(t_{n-1})\le x_0 <C(t_{n})$, where $t_{n-1}$ and $t_n$ are two successive time instants out of the $\rm N_{pdf}$ sample points at which $P(t)$ is computed, then the neutrino arrival time $t_a$ corresponding to $x_0$ is given by
\begin{equation}
    t_a=t_{ n-1} +\left[\frac{t_{ n}-t_{{ n}-1}}{C(t_{ n})-C(t_{ n-1})}\right]\left(x_0-C(t_{ n-1})\right),
\end{equation}
where we have used linear interpolation to compute $C^{-1}(x_0)$. 

\subsection{Neutrino event binning and window function}
\label{bins}
The $\rm N_{\nu}$ neutrino arrival times (generated as described in Section \ref{arrival_times_subsub}) are distributed into a total of $\rm N_{\rm bins}$ uniform time bins between $t_i$ and $t_f$. The mid-point of each time bin is chosen to be the representative time corresponding to that bin. The $k^{\rm th}$ bin with representative time $t_k$ contains $T(t_k)$ number of neutrinos, where $k=0,1,2,...,\rm N_{\rm bins}-1$. Here, $T$ is the function that maps the representative time of the $k^{\rm th}$ bin to the number of neutrinos in the $k^{\rm th}$ bin. For neutrino luminosity decay coefficient $\alpha>0$ (see equation \ref{PDF}), the amplitude of PDF $P(t)$ and hence the amplitude of neutrino counts function $T$ decreases as a function of time. This power law decay results in spurious peaks near zero frequency in the Fourier power spectrum (described in Section \ref{search_algos}), because the values of $P(t)$ at $t=t_i$ and $t=t_f$ do not match. The values at the edges of $P(t)$ have to match because the Discrete Fourier Transform (DFT) assumes that the function remains periodic outside the domain of the function that is being transformed. To suppress the spurious peaks, we multiply the function $T$ with a window function $w(t)$. We choose the function $w(t)$ such that it is zero at $t=t_i$ and $t=t_f$. We find that a Hann window $w(t)=\sin^2\left(\frac{\pi(t-t_0)}{t_{\rm \left[N_{\rm bins}-1\right]}-t_0}\right)$ is sufficient for our purposes. Hence the neutrino counts function after multiplication with the window function is,
\begin{equation}
    \label{neutrino_func}
    T_w(t)=T(t)\sin^2\left(\frac{\pi(t-t_0)}{t_{\rm \left[N_{\rm bins}-1\right]}-t_0}\right).
\end{equation}

\subsection{Search algorithms for periodic signals}
\label{search_algos}
To extract the frequency content in the neutrino counts data, we perform the Discrete Fourier Transform. The expression for the DFT with one frequency parameter is
\begin{equation}
\label{1d_DFT}
    D_{1}(\omega)=\sum_{k=0}^{\rm N_{\rm bins}-1} T_w(t_k)e^{-i\omega (t_k-t_0)}.
\end{equation}
The corresponding power spectrum is
 \begin{equation}
 \label{1d_power}
     P_1(\omega)=|D_1(\omega)|^2.
 \end{equation}
 The power spectrum $P_1(\omega)$ is sufficient to extract the frequency content of periodic signals with a constant period. We provide a simple intuitive explanation of the technique. We can expand $T_w(t)$ as a Fourier series as follows:
 \begin{equation}
 \label{fourier}
     T_w(t)=\sum_{n=-\infty}^{\infty} c_ne^{in\omega_{\rm p}(t-t_0)}.
 \end{equation}
Here $\omega_{\rm p}$ is the angular frequency of the periodic component in $T(t)$. From equations \ref{1d_DFT} and \ref{fourier},
\begin{equation}
\label{prod}
    D_1(\omega)=\sum_{k=0}^{\rm N_{\rm bins}-1}\sum_{n=-\infty}^{\infty}c_ne^{in\omega_{\rm p}(t_k-t_0)}e^{-i\omega (t_k-t_0)}.
\end{equation}
It is straightforward to observe that peaks in $P_1(\omega)=|D_1(\omega)|^2$ occur when the complex exponentials in the summation in equation \ref{prod} are in phase, which occurs at $\omega=n\omega_{\rm p}$ for integer $n$. The power spectrum also contains peaks corresponding to the window function $w(t)$, but these can be ignored because the frequency of $w(t)$ is much smaller than the PNS rotation frequencies we consider in this paper. For the modulating functions used in this paper (see Table \ref{mod_func_table}), the largest peak of $P_1(\omega)$ occurs at $\omega=\omega_{\rm p}$, that is at $n=1$. The peaks corresponding to $n=0$ are not relevant because we do not include zero frequency in the frequency search space. We note that it is possible to construct modulating functions such that the largest peak of the power spectrum occurs at $n>1$. If the PNS emits neutrinos with a modulating function that results in the largest peak occurring at $n>1$, then this has to be taken into account when determining the angular frequency of rotation of the PNS.

In an unlikely case, if there are multiple exactly identical equally spaced hot-spots on the PNS, then the Fourier power spectrum yields some multiple of the actual frequency of rotation of the PNS. For example, the case where there are two exactly identical hot-spots on opposite sides of the PNS resulting in a modulation as shown in Figure \ref{pdf_t3s} is exactly similar to the case where there is only one hot-spot but the PNS rotates twice as fast. But the probability of such a symmetry existing is small. On the other hand, if there are multiple asymmetric hot-spots on the PNS as shown in the right panel of Figure \ref{pdf_t3s_B2e15}, we can extract the frequency of rotation of the PNS from the neutrino counts data.  

To model spindown of PNSs, we extend the DFT to three parameters. From the simulations in \cite{Prasanna2022}, we find that PNS angular frequency as a function of time can be modelled as
 \begin{equation}
 \label{omega_model}
     \Omega(t)=\omega_1+\omega_2e^{-f_3(t-t_0)},
 \end{equation}
 where $\omega_1$, $\omega_2$ and $f_3$ are constants. Since the angular frequency is now a function of time, we use an integral of the angular frequency in the exponent of the DFT expression as follows:
 \begin{align}     
     D_3(\omega_1,\omega_2,f_3)&=\sum_{k=0}^{\rm N_{\rm bins}-1} T_w(t_k)e^{-i\int_{t_0}^{t_k}\Omega(t)dt}\\
     &=\sum_{k=0}^{\rm N_{\rm bins}-1} T_w(t_k)e^{-i\left[\omega_1(t_k-t_0)-\frac{\omega_2}{f_3}e^{-f_3(t_k-t_0)}+\frac{\omega_2}{f_3}\right]}.
 \end{align}
Since we are interested in the power spectrum only, we can ignore the constant phase terms in the exponent. Hence we get,
\begin{equation}
    \label{3d_DFT}
    D_3(\omega_1,\omega_2,f_3)=\sum_{k=0}^{\rm N_{\rm bins}-1} T_w(t_k)e^{-i\left[\omega_1(t_k-t_0)-\frac{\omega_2}{f_3}e^{-f_3(t_k-t_0)}\right]}.
\end{equation}
The corresponding power spectrum is
\begin{equation}
    \label{3d_power}
    P_3(\omega_1,\omega_2,f_3)=|D_3(\omega_1,\omega_2,f_3)|^2.
\end{equation}

The frequencies at which peaks occur in the power spectrum correspond to the input signal frequency. We provide an intuitive explanation of this technique. For a neutrino counts function with a non-constant angular frequency, we can think of the Fourier series as follows:
\begin{equation}
\label{mod_fourier}
    T_w(t)=\sum_{n=-\infty}^{\infty} c_ne^{ing(t-t_0)},
\end{equation}
where $g(t-t_0)$ is the phase function of $T_w(t)$. We can think of $g(t-t_0)$ as a function in the function space with $1,t,t^2,t^3,.....$ as the bases. Hence, we can write $g(t-t_0)$ as
\begin{equation}
    g(t-t_0)=\sum_{m=0}^{\infty}a_{m}(t-t_0)^m,
\end{equation}
where $a_m$ is a constant. The coefficients $a_m$ for $m\geq2$ are proportional to the successive time derivatives of the PNS angular frequency of rotation. For the case of $a_{m}=0$ for $m=0$ and all $m>1$, the above expression reduces to the regular exponential Fourier series (equation \ref{fourier}). Using this expansion in equation \ref{3d_DFT}, 
\begin{equation}
\begin{split}
 D_3(\omega_1,\omega_2,f_3)= \sum_{k=0}^{\rm N_{\rm bins}-1}  \sum_{n=-\infty}^{\infty}   c_ne^{ing(t_k-t_0)}  \\
 \times e^{-i\left[\omega_1(t_k-t_0)-\frac{\omega_2}{f_3}e^{-f_3(t_k-t_0)}\right]}.
\end{split}
\end{equation}
Peaks occur in the power spectrum $P_3(\omega_1,\omega_2,f_3)$ when
\begin{equation}
    ng(t_k-t_0)=\omega_1(t_k-t_0)-\frac{\omega_2}{f_3}e^{-f_3(t_k-t_0)}.
\end{equation}
Expanding the exponential as a series, we get,
\begin{equation}
\begin{split}
    n\sum_{m=0}^{\infty}a_{m}(t_k-t_0)^m=-\frac{\omega_2}{f_3}+(\omega_1+\omega_2)(t_k-t_0)\\
    -\frac{\omega_2}{2f_3}\left[f_3(t_k-t_0)\right]^2-\sum_{m=3}^{\infty}\frac{\omega_2}{m!f_3}\left[-f_3(t_k-t_0)\right]^m.
    \end{split}
\end{equation}
From the above equation, the values of $\omega_1$, $\omega_2$ and $f_3$ at the peak of the power spectrum are specified for each $n$. For the modulating functions considered in this paper (see Table \ref{mod_func_table}), the largest peak occurs at $n=1$. Peaks corresponding to $n=0$ are not relevant because we do not include zero frequency in the search space.

As described in Section \ref{arrival_times_subsub}, we use pseudo-random numbers to generate the neutrino arrival times. We find that the maximum power in the Fourier power spectrum depends on the random seed. The variation in maximum power is due to the fluctuation in the binned neutrino counts caused by different random seeds of the pseudo-random number generator. To get an idea of what we can expect from the next Galactic supernova, we repeat the procedure of computing the power spectrum 250 times by using 250 different random seeds to generate the neutrino arrival times and compute the power spectrum for each. For each random seed, we note the properties of the largest peak in the power spectrum and present the median of the results in this paper.

\subsection{Determining the False Alarm Rate (FAR)}
\label{far_sub}
To determine the uncertainty in the power spectrum results, we use the False Alarm Rate (FAR). Under the assumption that a signal (the neutrino counts function in this case) does not contain a periodic modulation, FAR is the probability that such a signal produces a maximum power in the power spectrum that is greater than or equal to the maximum power in the power spectrum of the signal with periodic modulation. That is, we use FAR to determine if the peak is due to actual periodicity or just due to a random fluctuation. We calculate FAR using a distribution of maximum power in the Fourier power spectrum in the absence of a periodic signal in the neutrino counts function. In equation \ref{PDF}, we have a periodic component and a non-periodic component, both of which decay as a function of time. Hence, the neutrino counts function also contains a periodic and a non-periodic part. We use the neutrino counts data $T(t)$ before multiplication with the window function and fit a power law and obtain the power law exponent $\alpha_{\rm b}$. Then the probability distribution without the periodic modulation is
\begin{equation}
\label{bootPDF}
    P_{\rm b}(t)=A_{\rm b}\left(\frac{t_i}{t}\right)^{\alpha_{\rm b}}.
\end{equation}
Enforcing $\int_{t_i}^{t_f}P_{\rm b}(t)dt=1$ fixes the the value of $A_{\rm b}$:
\begin{equation}
    A_{\rm b}=\frac{1-\alpha_{\rm b}}{t_i^{\alpha_{\rm b}}}\left[\frac{1}{t_f^{\alpha_{\rm b}-1}}-\frac{1}{t_i^{\alpha_{\rm b}-1}}\right]^{-1}.
\end{equation}

We again use the transformation method described in Section \ref{arrival_times_subsub} to draw random neutrino arrival times that follow the PDF $P_{\rm b}(t)$. We generate $\rm N_{\nu}$ neutrino arrival times and distribute these into a total of $\rm N_{\rm bins}$ bins. The neutrino counts thus generated just have a power law decay without the periodic modulation. We multiply the neutrino counts with the window function $w(t)$ and compute the power spectrum. To characterize FAR at a rate $r$, about $10/r$ power spectra without the periodic modulation need to be computed \citep{VanderPlas2018}. We generate a total of $\rm N_{\rm b}=1000$ power spectra without the periodic modulation in this paper to characterize FAR within $r=1\%$. To compute each of the $\rm N_{\rm b}$ power spectra, we generate the neutrino arrival times following $P_{\rm b}(t)$ with different random seeds. We name the set of $\rm N_b$ power spectra as the `non-periodic set'. Out of the $\rm N_{\rm b}$ power spectra of the non-periodic set, if the maximum power exceeds the maximum power in the power spectrum of the neutrino time series with periodic modulation $n$ times, then the false alarm rate is given by FAR$=n/\rm N_{\rm b}$.

\section{Results}
\label{results}
\begin{figure*}
\centering{}
\includegraphics[width=\textwidth]{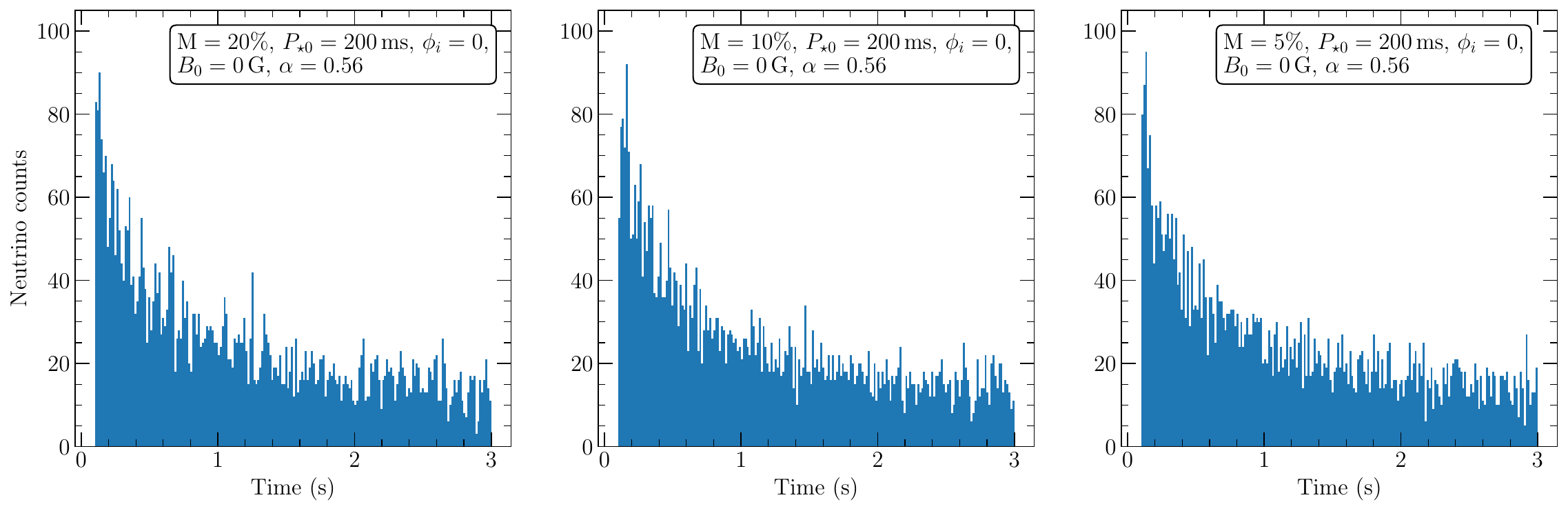}
\caption{Histogram of neutrino counts with a $\sin^2$ modulation at a constant PNS spin period of 200\,ms for different values of periodic content M with a total of $5\times 10^{3}$ neutrinos distributed into 200 bins. We use the label polar magnetic field strength $B_0=0$\,G to imply a constant PNS spin period (refer to the caption of Figure \ref{pdf_t3s}).} 
\label{counts_5e3_t3s}
\end{figure*}

\begin{figure*}
\centering{}
\includegraphics[width=\textwidth]{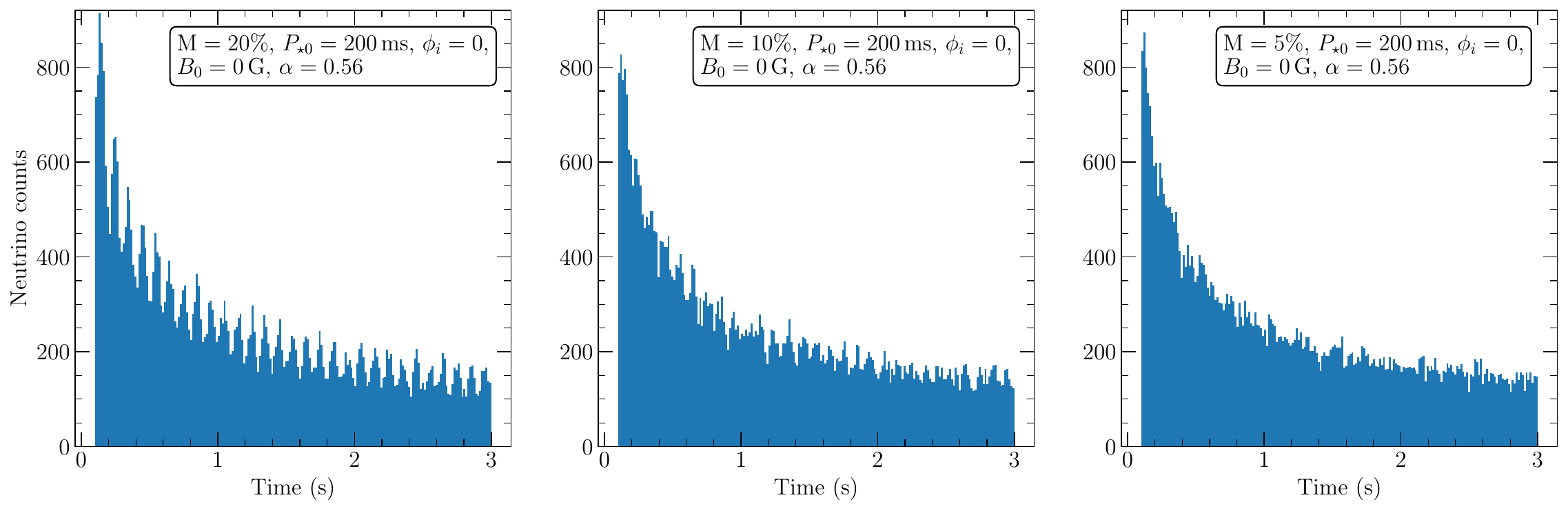}
\caption{Same as Figure \ref{counts_5e3_t3s}, but with a total of $5\times 10^{4}$ neutrinos.} 
\label{counts_5e4_t3s}
\end{figure*}

The goal of this paper is to explore the possibility of determining the rotation rate and spindown (if there is any) of a PNS from the neutrino counts as a function of time. Figures \ref{counts_5e3_t3s} and \ref{counts_5e4_t3s} show the histogram of neutrino counts (generated using the method described in Section \ref{arrival_times_subsub}) with a $\sin^2$ modulation distributed into 200 bins for a total of $5\times10^{3}$ and $5\times 10^{4}$ neutrinos respectively for various values of periodic content M at a constant PNS spin period of 200\,ms. The histograms clearly show the presence of periodic modulation in neutrino counts as a function of time. The histograms use 200 bins just for the purpose of visualization.

We apply the Fourier transform techniques to the binned neutrino counts to extract the frequency content of the neutrino signal as described in Section \ref{search_algos}. We first present results for PNSs rotating with roughly a constant spin period during the cooling phase. This can occur if the surface  magnetic field strength of the PNS is $\lesssim 10^{14}$\,G resulting in long spindown timescales compared to the PNS cooling timescale and hence negligible spindown \citep{Prasanna2022, Prasanna2023}. Negligible spindown during the cooling phase can also occur at high surface magnetic fields $\sim 10^{15}$\,G if the PNS is born rapidly rotating with initial spin period $P_{\star 0}\lesssim 20$\,ms, which again results in long spindown timescales compared to the cooling timescale \citep{Prasanna2023}. For PNSs with roughly constant rotation rates during the cooling phase, a one-parameter DFT (see equations \ref{1d_DFT} and \ref{1d_power}) is sufficient to determine the spin period of the PNS. We label the constant PNS spin period cases with polar magnetic field strength $B_0=0$\,G, because spindown is negligible when the magnetic field strength is small as mentioned above.

\subsection{Cases without spindown}
\label{const_spin}
Table \ref{table1} shows the properties of the largest peak in the Fourier power spectrum for a $\sin^2$ modulation (see Figure \ref{pdf_t3s}) as a function of spin period $P_{\star 0}$ of the PNS, periodic content M, initial phase of the neutrino signal, and total number of neutrinos. Most of the results presented are with an initial phase $\phi_i=0$. We present some results with various values of $\phi_i>0$ to show that detection is not sensitive to the initial phase. We describe the quantities reported in Table \ref{table1} in detail. The number of PDF $P(t)$ samples $\rm N_{\rm pdf}$ used to generate the neutrino counts has to be sufficiently large to accurately represent the PDF. $\rm N_{\rm pdf}$ is determined by the spin period $P_{\star0}$ and the duration of the neutrino signal. As the PNS spin period decreases, a larger value of $\rm N_{\rm pdf}$ is required to accurately represent the PDF. For PNS spin periods $P_{\star 0}\ge 10$\,ms, we find that $2\times 10^{4}$ PDF samples are sufficient. For $P_{\star 0}=5$\,ms, we use $4\times 10^{4}$ PDF samples and for $P_{\star 0}=1$\,ms, we use $2\times 10^{5}$ PDF samples. Before settling on a specific value of $\rm N_{\rm pdf}$, we have run simulations with a larger number of samples and confirmed that the results do not change (see Table \ref{table1}). Using a smaller value of $\rm N_{pdf}$ than the above mentioned values leads to underestimation of the maximum power in the power spectrum and this results in a larger value of FAR. 

The frequency grid to compute the Fourier power spectrum spans the the whole range of possible PNS frequencies of rotation at birth. The angular frequencies of the search space ranges from $2\pi$\,rad s$^{-1}$ (corresponding to periodicity of 1\,s) to $5000\pi$\,rad s$^{-1}$ (corresponding to periodicity of 0.4\,ms). We compute the power spectrum at $4\times 10^{4}$ equally spaced frequencies on the grid. The general recommendation is to use a total of $n_0(t_f-t_i)f_{\max}$ points on the frequency grid with $n_0\sim5$, where $f_{\max}$ is the maximum frequency on the grid \citep{VanderPlas2018}. The number of time bins $\rm N_{bins}$ should satisfy the Nyquist sampling criterion which states that the sampling frequency in the time domain should be larger than twice the bandwidth of the signal to correctly extract the frequency content of the signal. Since the maximum frequency in our search space is 2500\,Hz, we require at least 14,500 time bins for $t_i=0.1$\,s and $t_f=3$\,s (the reason for choosing these values of $t_i$ and $t_f$ is mentioned in the next paragraph) to satisfy the Nyquist criterion. $\rm N_{bins}$ also depends on the PNS spin period, which is related to the periodicity of the neutrino signal. There should be enough number of bins per period of the input neutrino signal to correctly represent the input signal. We find that using too few bins results in a smaller value of the maximum power in the power spectrum and hence a larger false alarm rate. We find from our simulations that $\rm N_{bins}=16000$ is sufficient for initial PNS spin periods $\ge 5$\,ms. For $P_{\star0}=1$\,ms, we find that 64000 time bins are required. Before settling on a particular value of $\rm N_{bins}$, we have run simulations with a larger number of bins to ensure that the results do not change (see Table \ref{table1}). We emphasize that detection is possible as long as the Nyquist condition is satisfied, but using a smaller number of bins than those mentioned above leads to overestimation of FAR. In the event of a Galactic supernova, the number of bins has to be set as mentioned above based on the maximum frequency in the search space to correctly estimate the FAR.

As the duration of the neutrino signal increases, more time bins are required to satisfy the Nyquist sampling criterion. Also, more PDF samples need to be generated and the Fourier power spectrum has to be computed at a larger number of points. Since we cover a large parameter space, to reduce the computational complexity, we limit the neutrino signal duration to 2.9\,s. However, in the event of a Galactic supernova, we can utilise the entire neutrino signal. We set $t_i=0.1$\,s and $t_f=3$\,s as the times of arrival of the first and last neutrinos respectively. We do not set $t_i=0$\,s to ensure that the PDF $P(t)$ is finite (see equation \ref{PDF}). A non-zero time of arrival of the first neutrino is also required to quantify the significance of the Fourier power spectrum peaks (described in Section \ref{far_sub}). For a real neutrino signal from a supernova, if we consider the time of arrival of the first neutrino as $t=0$\,s, $t_i>0$\,s means that we ignore the neutrinos that arrived before $t_i$ for our analysis. We note that the neutrino signal in the first few seconds can be dominated by the explosion physics rather than the PNS cooling \citep{Li2023}. For such a scenario, $t=0$\,s in this paper corresponds to the time of arrival of the neutrino after the PNS cooling phase begins. To show that PNS rotation can be detected even with $t_i<0.1$\,s, we present results from a simulation with $t_i=0.01$\,s (see Table \ref{table1}). In a different scenario, it is possible that the periodicity in the neutrino signal does not exist for some time after the arrival of the first neutrino (this could occur if there is a delay in formation of neutrino emission hot-spots). In such a case, we just have to discard the part of the signal without periodicity. We show one result with $t_i=1.0$\,s and $t_f=3.9$\,s to show that detection is not sensitive to the start time (see Section \ref{spindown_results} and Table \ref{table2}).

As described in Section \ref{search_algos}, we generate 250 sets of neutrino counts with different random seeds for each set of input parameters. Each set of neutrino counts produces a different Fourier power spectrum. It is possible that a fraction of the 250 sets yields maximum power in the power spectrum at an `incorrect' frequency. We define a detection as `incorrect' if the largest peak in the power spectrum occurs at a frequency that is not within 5\% of the input signal frequency. We report the fraction of incorrect detections $F$ in Table \ref{table1}. Since the results shown in Table \ref{table1} are with a $\sin^2$ modulation, the input neutrino signal frequency is twice the frequency of rotation of the PNS. In case of an incorrect detection, we set the value of power in the Fourier power spectrum to zero and the associated FAR is set to 100\%. In Table \ref{table1}, $\omega_{\rm p}$ is the median of correctly detected angular frequencies. $E$ is the median of the deviation of the detected angular frequencies from the input angular frequency. $\alpha_{\rm b}$ is the mean of the power law exponents obtained from a power law fit to each of the 250 sets of neutrino counts. 

We use the value of $\alpha_{\rm b}$ to obtain the $\rm N_{b}$ power spectra without the periodic modulation to determine the false alarm rate as described in Section \ref{far_sub}. We use the $\rm N_{b}$ power spectra to obtain 250 values of the false alarm rate corresponding to the 250 sets of neutrino counts with periodic modulation. FAR reported in Table \ref{table1} is the median of these 250 values of false alarm rate. All the results in Table \ref{table1} are with $\rm N_b=1000$. We have run a simulation with $P_{\star0}=200$\,ms and $\rm M=13.5\%$ with $\rm N_b=4000$ and obtain $\rm FAR= 1.65\%$ which is close to $\rm FAR=1.66\%$ obtained with $\rm N_b=1000$. 

$\rm FAR=2\%$ corresponds to $2\sigma$. Hence, we require $\rm FAR\le 2\%$ for a detection to be statistically significant. Ideally, to determine the median FAR, we have to generate $\rm N_b$ power spectra without the periodic modulation (that is the non-periodic set, see Section \ref{far_sub}) for each of the 250 different sets of neutrino counts because each set gives a different power law exponent, but this is computationally expensive. Hence, we generate a total of 11 non-periodic sets. 10 of the 11 non-periodic sets are with power law exponents ranging from 0.5 to 0.6 obtained from 10 different sets of neutrino counts with periodic modulation and the eleventh non-periodic set is with $\alpha_{\rm b}$ which is the average power law exponent obtained from 250 sets of neutrino counts. For each of the 11 different power law exponents, we generate $\rm N_b$ sets of neutrino counts following the PDF $P_{\rm b}(t)$ (equation \ref{bootPDF}) and obtain the `non-periodic' power spectrum. For each non-periodic set, we obtain 250 different values of FAR corresponding to the 250 sets of neutrino counts with periodic modulation. We find that the median FAR obtained from each of the 11 non-periodic sets agree with each other. However, we note that using the average $\alpha_{\rm b}$ which is close to the value of $\alpha$ used in equation \ref{PDF} gives the best estimate of FAR because it closely follows the power law decay of the original periodic signal and hence we report FAR obtained using the average value $\alpha_{\rm b}$.

Figure \ref{FAR_1d_fig} shows the false alarm rate obtained from 1-parameter DFT as a function of periodic content M and initial PNS spin period $P_{\star0}$. Since the polar magnetic field strength $B_0=0$\,G in these simulations, the PNS spin period remains constant as a function of time and hence 1-parameter DFT is sufficient. We find that FAR is less than 1\% for all values of $P_{\star0}$ with a total of $\rm N_{\nu}=5\times 10^{4}$ neutrinos for $\rm M\geq 5\%$. With $\rm N_{\nu}=5\times 10^{3}$, all PNS spin periods can be detected with $\rm FAR <2\%$ if $\rm M\gtrsim 13\%$. As the total number of neutrinos increase, the signal strength increases and detection is possible for lower values of M. 

Figure \ref{1d_DFT_5e3} shows the normalized one-parameter Fourier power spectrum (see equations \ref{1d_DFT} and \ref{1d_power}) at a constant PNS spin period of 200\,ms for various values of periodic content M with $5\times 10^{3}$ neutrinos for a $\sin^2$ modulation. The largest peak occurs at $\omega= 62.81$\,rad s$^{-1}$ for $\rm M=20\%$. The angular frequency at the peak is twice the PNS angular frequency of rotation because we use a modulating function proportional to $\sin^2$. For $\rm M=10\%$ and 5\%, there are peaks at $\omega= 62.81$\,rad s$^{-1}$, but there are larger peaks at other values of $\omega$. Figure \ref{1d_DFT_5e4} shows that the largest peak occurs at $\omega= 62.81$\,rad s$^{-1}$ with $5\times 10^{4}$ neutrinos for $\rm M= 20\%, \ 10\%$ and 5\%. This is because as the periodic content increases and/or as the number of neutrinos increase, the signal strength and hence the magnitude of power at the input signal frequency increases.

\subsection{Analytic estimate of FAR}
\label{analytic}

We provide analytic estimates of the false alarm rate (FAR) to compare with our results. The Baluev method provides an analytic upper limit for the FAR \citep{Baluev2008}. If $z$ is the maximum power in the normalized Fourier power spectrum, then the upper limit of the FAR is:
\begin{equation}
    {\rm FAR_{b}} = 1- \left[1-{\rm e}^{-z}\right]{\rm e}^{-\tau(z)},
\end{equation}
where $\tau(z)=(t_f-t_i)f_{\rm max}\sqrt{z}{\rm e}^{-z}$ with $f_{\rm max}$ being the maximum frequency on the grid. The above expression assumes a power spectrum normalization $P_1(\omega)=(1/{\rm N_{bins}}) |D_1(\omega)|^2$ which differs from the normalization used in this paper by a factor of $1/{\rm N_{bins}}$ (see equation \ref{1d_power}). We note that the normalization of the power spectrum does not affect the results because we use the same normalization to compute the power spectrum of the periodic signal and the pure power law signal used to compute the FAR. To compute the upper limit of the FAR, we first compute the 1-parameter Fourier power spectrum for the function $u(t)=P(t){\rm N_{\nu}}(t_f-t_i)/{\rm N_{bins}}$, where $P(t)$ is the PDF from equation \ref{PDF}. $u(t)$ gives the neutrino counts that exactly follows the PDF $P(t)$. The neutrino counts generated using the uniform random numbers that we use for the analysis in this paper (see Section \ref{arrival_times_subsub}) have some fluctuations and hence do not exactly follow the PDF $P(t)$.

\begin{figure}
\centering{}
\includegraphics[width=0.87\linewidth]{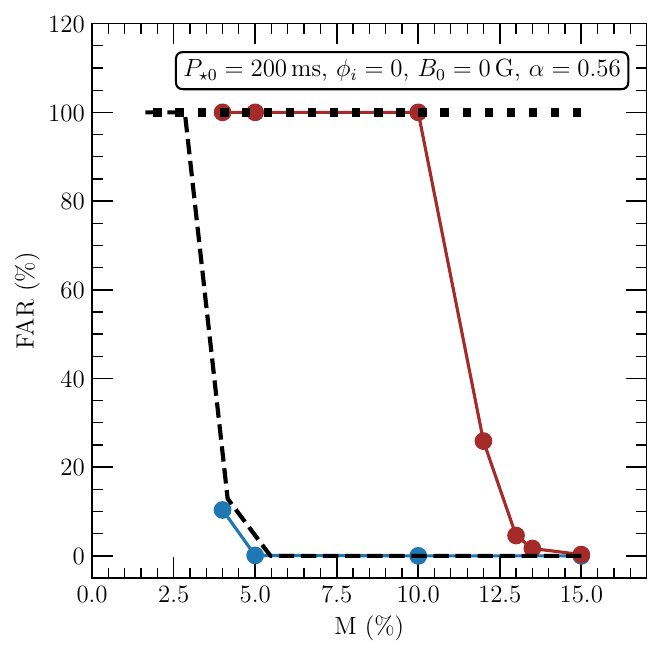}
\caption{False alarm rate (FAR) as a function of periodic signal content at a constant PNS spin period $P_{\star0}=200$\,ms. The solid blue and brown lines show the FAR obtained using the 1-parameter DFT (see Table \ref{table1}) for $5\times 10^{4}$ and $5\times 10^{3}$ neutrinos respectively. The dashed and dotted black lines show the estimate of FAR obtained using the Baluev method (see Section \ref{analytic}).} 
\label{FAR_Baluev}
\end{figure}

Figure \ref{FAR_Baluev} shows the FAR as a function of periodic content M for a constant PNS spin period of 200\,ms. The solid blue and brown lines show the 1-parameter DFT results from Table \ref{table1} for $\rm N_{\nu}=5\times 10^{4}$ and $5\times 10^{3}$ respectively. The dashed and dotted black lines show the upper limit of the FAR obtained by applying the Baluev method to the normalized power spectrum of $u(t)$ for $5\times 10^{4}$ and $5\times 10^{3}$ neutrinos respectively. For $\rm N_{\nu}=5\times 10^{3}$, the Baluev FAR estimate is 100\% for all values M, which does not provide a useful comparison. But, the Baluev FAR estimate for $\rm N_{\nu}=5\times 10^{4}$ is close to our results, which shows that the method used to obtain the FAR in this paper gives reasonable results.

\subsection{Cases with spindown}
\label{spindown_results}
Table \ref{table2} shows the results from the 3-parameter DFT (see equations \ref{3d_DFT} and \ref{3d_power}). We use the 3-parameter DFT to extract information about PNSs that undergo significant change in spin period during the cooling phase. PNSs born rotating `slowly' with initial spin period $P_{\star0}\gtrsim 100$\,ms and polar magnetic field strength $B_0\gtrsim 10^{15}$\,G can spindown to periods greater than 1\,s during the cooling phase that lasts $\sim 1-100$\,s \citep{Prasanna2022}. The search space for the 3-parameter DFT spans $2\pi$\,rad s$^{-1}\le \omega_1\le 100\pi$\,rad s$^{-1}$, $2\pi$\,rad s$^{-1}\le \omega_2\le 100\pi$\,rad s$^{-1}$, and $-0.1$\,s$^{-1}\le f_3\le 2$\,s$^{-1}$. The search space is chosen based on the results from \cite{Prasanna2022} and \cite{Prasanna2023}. PNSs with initial spin period $P_{\star0}\lesssim 20$\,ms are unlikely to spin down significantly during the cooling phase because the spindown timescales are much larger than the PNS cooling timescale \citep{Prasanna2023}, which is why we exclude PNS spin periods $< 20$\,ms from the search space. We restrict the frequency grid due to computational complexity as well. As we include higher frequencies on the grid, $\rm N_{bins}$ increases, leading to a significant increase in computation time. We emphasize that if a Galactic supernova results in a PNS with order of milliseconds initial spin period that spins down significantly, we can expand the search space and detect the spindown of such PNSs as well.

We evaluate the 3-parameter Fourier power spectrum  at 400 equally spaced points on each of $\omega_1$ and $\omega_2$ axes and 128 equally spaced points on the $f_3$ axis. As mentioned in the 1-parameter DFT case, the general recommendation is to have $n_0(t_f-t_i)f_{\rm max}$ samples on the grid, where $f_{\rm max}$ is the maximum frequency on the grid, with $n_0\sim5$. In the grid chosen for the 3-parameter DFT, we have $n_0=2.76$ for the $\omega_1$ and $\omega_2$ axes and $n_0=22$ for the $f_3$ axis. Such a grid has been chosen to accurately determine the spindown parameter $f_3$. With these choices, we have sufficiently over-sampled the grid so as to not miss any peak. To test this, we have run one simulation with $P_{\star0}=200$\,ms and $B_0=2\times 10^{15}$\,G for a $\sin^2$ modulation with 800 grid points each on the $\omega_1$ and $\omega_2$ axes and 32 grid points on the $f_3$ axis which results in $n_0=5.52$ along all the three axes. We find that these results match with the other grid scheme (see Table \ref{table2}).

We find that $2\times 10^{4}$ PDF samples are adequate to correctly represent $P(t)$. $\rm N_{bins}=300$ is sufficient to satisfy the Nyquist sampling criterion in the individual $\omega_1$, $\omega_2$ and $f_3$ grids. But, we choose $\rm N_{bins}=800$ to make sure that there are a sufficient number of bins per input neutrino signal period. We have run a couple of simulations with 1600 bins and we find that the results do not change significantly (see Table \ref{table2}). At $P_{\star0}=50$\,ms, $B_0=10^{15}$\,G and $\rm M=4\%$, the values of FAR vary slightly as a function of number of bins. As the FAR exceeds $\sim4-5\%$, FAR is sensitive to even minor variations in the power spectrum caused by changing the number of bins, which is why the variation of FAR between 6\% and 14\% as a function of number of bins in this case is not concerning. We also note that conclusion about detection remains the same as we consider any $\rm FAR > 2\%$ to be consistent with noise. 

In Table \ref{table2}, $\omega_{\rm 1p}$, $\omega_{\rm 2p}$ and $f_{\rm 3p}$ are the values of $\omega_{1}$, $\omega_{2}$ and $f_3$ (see equation \ref{omega_model}) that result in minimum deviation from the angular frequency of the input signal out of the 250 different sets of neutrino counts with periodic modulation. For the case of non-constant PNS angular frequency, we define the error or deviation from the angular frequency of the input signal as follows:
\begin{equation}
    {\rm deviation}=\frac{1}{N}\sum_{t} \frac{|\Omega_{\rm d}(t)-\Omega(t)|}{\Omega(t)},
\end{equation}
where $N$ is the total number of time instances at which the angular frequencies are evaluated, $\Omega(t)$ is the angular frequency of the input neutrino signal, and $\Omega_{\rm d}(t)=\omega_{\rm 1d}+\omega_{\rm 2d}e^{-f_{\rm 3d}(t-t_0)}$ is the angular frequency obtained from the largest peak parameters of the Fourier power spectrum. 

In Table \ref{table2}, we present a few results with neutrino luminosity decay coefficient $\alpha=0.8$. We find that the results are similar to the results obtained with $\alpha=0.56$. In the event of a supernova, we can expect PNS cooling with $\alpha$ evolving as a function of time. Hence, we present results from a simulation with hypothetical time dependent $\alpha$. For this simulation, we set a constant $\alpha=0.56$ for $t_i\le t\le 0.5$\,s and set $\alpha(t)=0.56+0.1(t-0.5\,{\rm s})$ for $0.5\,{\rm s}<t\le3$\,s. Even with such a model, we find that we can detect spindown and the results are similar to those with constant values of $\alpha$. 

We also present results from simulations with non-sinusoidal modulating functions. We show that spindown can be detected with a square wave modulation as shown in the middle panel of Figure \ref{pdf_t3s_B2e15} and a modulating function obtained from a combination of sinusoidal and square waves as shown in the right panel of Figure \ref{pdf_t3s_B2e15}. We emphasize that our detection algorithm and results are not limited to a particular choice of the modulating function. 

In Table \ref{table2}, we show one simulation at $P_{\star0}=200$\,ms, $B_0=2\times 10^{15}$\,G and a modulating function obtained from a combination of sinusoidal and square waves with $t_i=1.0$\,s and $t_f=3.9$\,s. We find that these results are close to the results obtained using $t_i=0.1$\,s and $t_f=3.0$\,s. This shows that the detection and FAR are not sensitive to the start time of the neutrino signal. Table \ref{table2} also shows results from a physically motivated modulating function as shown in Figure \ref{pdf_t3s_Chrismod} obtained assuming a small hot-spot (size $\ll$ PNS neutrinosphere radius) on the equator of the PNS.    

For all the results in Table \ref{table2}, we use $\rm N_b=1000$ non-periodic power spectra to determine the FAR (see Section \ref{far_sub} for details). We have run one simulation with $P_{\star0}=200$\,ms, $B_0=0$\,G and $\rm M=14\%$ with $\rm N_b=2000$. We find that $\rm FAR=0.78\%$ with $\rm N_b=2000$ is close to $\rm FAR=0.24\%$ with $\rm N_b=1000$. The small variation is consistent with statistical fluctuation.

\begin{figure}
\centering{}
\includegraphics[width=\linewidth]{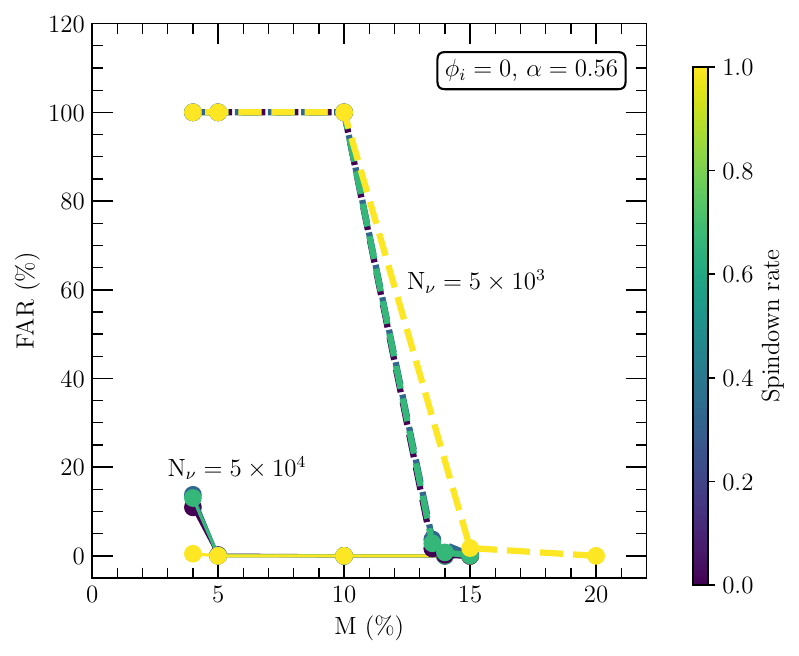}
\caption{False alarm rate (FAR) as a function of periodic signal content and spindown rate. Yellow (spindown rate of 1) corresponds to the fastest spindown case in Table \ref{table2} with PNS initial spin period $P_{\star0}=400$\,ms and $B_0=4\times 10^{15}$\,G. Dark violet with spindown rate of 0 corresponds to the case of constant PNS spin period in Table \ref{table2} with $P_{\star0}=200$\,ms and $B_0=0$\,G. The other two colors correspond to slow spindown ($P_{\star0}=50$\,ms and $B_0=10^{15}$\,G) and moderate spindown ($P_{\star0}=200$\,ms and $B_0=2\times10^{15}$\,G). This figure shows the results obtained using the 3-parameter DFT. All the results in this figure are with a $\sin^2$ modulation. Refer to the caption of Figure \ref{pdf_t3s_B2e15} for the meaning of spindown rates labelled in terms of $B_0$ and $P_{\star0}$.}
\label{FAR_2d_fig}
\end{figure}

Figure \ref{FAR_2d_fig} shows FAR as a function of M and spindown rate. Spindown rate of 0 corresponds to the case of constant PNS spin period and spindown rate of 1 corresponds to the case of fastest spindown in Table \ref{table2} with $P_{\star0}=400$\,ms and $B_0=4\times10^{15}$\,G. Similar to the results from 1-parameter DFT, even in these cases, we find that detection of spindown with $\rm FAR<1\%$ is possible with $5\times 10^{4}$ neutrinos if $\rm M \geq 5\%$. With $5\times 10^{3}$ neutrinos, detection with $\rm FAR<2\%$ is possible if $\rm M\gtrsim 13-15\%$.

\begin{table*}
	\caption{Results from 1-parameter DFT. All the results in this Table are with a $\sin^2$ modulation as shown in Figure \ref{pdf_t3s}. Simulations run with a $\sin^2$ modulation result in the largest peak in the Fourier power spectrum at twice the input signal frequency. Refer to the table footnotes for the meaning of the quantities listed (Table \ref{table1} continues on the next page).}
	\label{table1}
      \begin{threeparttable}
	\begin{tabular}[width=\textwidth]{@{}cccccccccccccccc@{}} 
 
		\hline
		 $P_{\star0}  ^{\rm (a)}$ & $\Omega_{\star0} ^{(\rm b)}$ & $B_0  ^{(\rm c)}$ & M$ ^{(\rm d)}$ & $\phi_{i} ^{(\rm e)}$ & $\alpha ^{(\rm f)}$ & $\rm N_{\nu} ^{(\rm g)}$ & $\rm N_{\rm bins} ^{(\rm h)}$  & $t_i ^{(\rm i)}$ & $t_j ^{(\rm j)}$ & $\rm N_{\rm pdf} ^{(\rm k)}$  & $\alpha_{\rm b} ^{(\rm l)}$ & $\omega_{\rm p} ^{(\rm m)}$ & $F$ $ ^{(\rm n)}$ & $E$ $ ^{(\rm o)}$ &FAR$ ^{(\rm p)}$ \\ 
		(ms) &  (rad s$^{-1}$) & (G) & & & & & & (s) & (s) & & &   (rad s$^{-1}$) & & & \\
		\hline

\\

  200 & 31.42 & 0 & 20\%& 0 & 0.56 & $5\times 10^{3}$ & 16000  & 0.1 & 3.0 & $2 \times 10^{4}$ & 0.56 & 62.81  & 0 & 0.03\% & 0\%  \\
  & & 0 & 20\%& $\pi$/6 & 0.56 & $5\times 10^{3}$ & 16000  & 0.1 & 3.0 & $2 \times 10^{4}$ & 0.57 & 62.81  & 0 & 0.03\% & 0\%  \\
  & & 0 & 20\%& $\pi$/4 & 0.56 & $5\times 10^{3}$ & 16000  & 0.1 & 3.0 & $2 \times 10^{4}$ & 0.58 & 62.81  & 0 & 0.03\% & 0\%  \\
  & & 0 & 20\%& $\pi$/3 & 0.56 & $5\times 10^{3}$ & 16000  & 0.1 & 3.0 & $2 \times 10^{4}$ & 0.58 & 62.81  & 0 & 0.03\% & 0\%  \\
  & & 0 & 20\%& $\pi$/2 & 0.56 & $5\times 10^{3}$ & 16000  & 0.1 & 3.0 & $2 \times 10^{4}$ & 0.57 & 62.81  & 0 & 0.03\% & 0\%  \\

  & & 0 & 15\%& 0 & 0.56 & $5\times 10^{4}$ & 16000  & 0.1 & 3.0 & $2 \times 10^{4}$ & 0.56 & 62.81  & 0 & 0.03\% & 0\%  \\
  & & 0 & 15\%& $\pi/6$ & 0.56 & $5\times 10^{4}$ & 16000  & 0.1 & 3.0 & $2 \times 10^{4}$ & 0.57 & 62.81  & 0 & 0.03\% & 0\%  \\
  & & 0 & 15\%& $\pi/4$ & 0.56 & $5\times 10^{4}$ & 16000  & 0.1 & 3.0 & $2 \times 10^{4}$ & 0.57 & 62.81  & 0 & 0.03\% & 0\%  \\
  & & 0 & 15\%& $\pi/3$ & 0.56 & $5\times 10^{4}$ & 16000  & 0.1 & 3.0 & $2 \times 10^{4}$ & 0.57 & 62.81  & 0 & 0.03\% & 0\%  \\
  & & 0 & 15\%& $\pi/2$ & 0.56 & $5\times 10^{4}$ & 16000  & 0.1 & 3.0 & $2 \times 10^{4}$ & 0.56 & 62.81  & 0 & 0.03\% & 0\%  \\

  & & 0 & 15\%& 0 & 0.56 & $5\times 10^{3}$ & 16000  & 0.1 & 3.0 & $2 \times 10^{4}$ & 0.56 & 62.81  & 0.07 & 0.03\% & 0.29\%  \\
  & & 0 & 15\%& $\pi$/6 & 0.56 & $5\times 10^{3}$ & 16000  & 0.1 & 3.0 & $2 \times 10^{4}$ & 0.57 & 62.81  & 0.10 & 0.03\% & 0.19\%  \\
  & & 0 & 15\%& $\pi$/4 & 0.56 & $5\times 10^{3}$ & 16000  & 0.1 & 3.0 & $2 \times 10^{4}$ & 0.57 & 62.81  & 0.12 & 0.59\% & 0.29\%  \\
   & & 0 & 15\%& $\pi$/3 & 0.56 & $5\times 10^{3}$ & 16000  & 0.1 & 3.0 & $2 \times 10^{4}$ & 0.57 & 62.81  & 0.11 & 0.03\% & 0.10\%  \\
   & & 0 & 15\%& $\pi$/2 & 0.56 & $5\times 10^{3}$ & 16000  & 0.1 & 3.0 & $2 \times 10^{4}$ & 0.56 & 62.81  & 0.10 & 0.03\% & 0.10\%  \\
   & & 0 & 15\%& 0 & 0.56 & $5\times 10^{3}$ & 16000  & 0.1 & 3.0 & $4 \times 10^{4}$ & 0.56 & 62.81  & 0.07 & 0.03\% & 0.29\%  \\
   & & 0 & 15\%& 0 & 0.56 & $5\times 10^{3}$ & 20000  & 0.1 & 3.0 & $2 \times 10^{4}$ & 0.56 & 62.81  & 0.07 & 0.31\% & 0.15\%  \\
   & & 0 & 15\%& 0 & 0.56 & $5\times 10^{3}$ & 16000  & 0.01 & 3.0 & $2 \times 10^{4}$ & 0.54 & 62.81  & 0.13 & 0.03\% & 0.19\%  \\

& & 0 & 13.5\%& 0 & 0.56 & $5\times 10^{3}$ & 16000  & 0.1 & 3.0 & $2 \times 10^{4}$ & 0.56 & 62.81 & 0.24 & 0.59\% & 1.66\%  \\
& & 0 & 13\%& 0 & 0.56 & $5\times 10^{3}$ & 16000  & 0.1 & 3.0 & $2 \times 10^{4}$ & 0.56 & 62.81 & 0.28 & 0.59\% & 4.59\%  \\
& & 0 & 12\%& 0 & 0.56 & $5\times 10^{3}$ & 16000  & 0.1 & 3.0 & $2 \times 10^{4}$ & 0.56 & 62.81 & 0.41 & 0.66\% & 25.88\%  \\

   & & 0 & 10\%& 0 & 0.56 & $5\times 10^{4}$ & 16000  & 0.1 & 3.0 & $2 \times 10^{4}$ & 0.56 & 62.81  & 0 & 0.03\% & 0\%  \\
   & & 0 & 10\%& 0 & 0.56 & $5\times 10^{3}$ & 16000  & 0.1 & 3.0 & $2 \times 10^{4}$ & 0.56 & 62.81 & 0.71 & 5945.55\% & 100\%  \\ 

   & & 0 & 5\%& 0 & 0.56 & $5\times 10^{4}$ & 16000  & 0.1 & 3.0 & $2 \times 10^{4}$ & 0.56 & 62.81  & 0.05 & 0.03\% & 0.10\%  \\ 

   & & 0 & 4\%& 0 & 0.56 & $5\times 10^{4}$ & 16000  & 0.1 & 3.0 & $2 \times 10^{4}$ & 0.56 & 62.81  & 0.32 & 0.66\% & 10.35\%  \\ \\ 

   50 & 125.66 & 0 & 20\%& 0 & 0.56 & $5\times 10^{3}$ & 16000  & 0.1 & 3.0 & $2 \times 10^{4}$ & 0.56 & 251.24  & 0 & 0.04\% & 0\%  \\ 

    &  & 0 & 15\%& 0 & 0.56 & $5\times 10^{4}$ & 16000  & 0.1 & 3.0 & $2 \times 10^{4}$ & 0.56 & 251.24  & 0 & 0.04\% & 0\%  \\
    &  & 0 & 15\%& 0 & 0.56 & $5\times 10^{3}$ & 16000  & 0.1 & 3.0 & $2 \times 10^{4}$ & 0.56 & 251.24  & 0.08 & 0.12\% & 0.10\%  \\

    &  & 0 & 13.5\%& 0 & 0.56 & $5\times 10^{3}$ & 16000  & 0.1 & 3.0 & $2 \times 10^{4}$ & 0.56 & 251.24  & 0.20 & 0.12\% & 0.63\%  \\

    &  & 0 & 10\%& 0 & 0.56 & $5\times 10^{4}$ & 16000  & 0.1 & 3.0 & $2 \times 10^{4}$ & 0.56 & 251.24  & 0 & 0.04\% & 0\%  \\
    &  & 0 & 10\%& 0 & 0.56 & $5\times 10^{3}$ & 16000  & 0.1 & 3.0 & $2 \times 10^{4}$ & 0.56 & 251.24  & 0.64 & 1098.31\% & 100\%  \\ 

    &  & 0 & 5\%& 0 & 0.56 & $5\times 10^{4}$ & 16000  & 0.1 & 3.0 & $2 \times 10^{4}$ & 0.56 & 251.24  & 0.04 & 0.04\% & 0.10\%  \\ 

    &  & 0 & 4\%& 0 & 0.56 & $5\times 10^{4}$ & 16000  & 0.1 & 3.0 & $2 \times 10^{4}$ & 0.56 & 251.24  & 0.31 & 0.12\% & 10.99\%  \\ \\

    10 & 628.32 & 0 & 20\%& 0 & 0.56 & $5\times 10^{3}$ & 16000  & 0.1 & 3.0 & $2 \times 10^{4}$ & 0.56 & 1256.56  & 0 & 0.01\% & 0\%  \\ 

     &  & 0 & 15\%& 0 & 0.56 & $5\times 10^{4}$ & 16000  & 0.1 & 3.0 & $2 \times 10^{4}$ & 0.56 & 1256.56  & 0 & 0.01\% & 0\%  \\
     &  & 0 & 15\%& 0 & 0.56 & $5\times 10^{3}$ & 16000  & 0.1 & 3.0 & $2 \times 10^{4}$ & 0.56 & 1256.56  & 0.09 & 0.02\% & 0\%  \\

     &  & 0 & 13.5\%& 0 & 0.56 & $5\times 10^{3}$ & 16000  & 0.1 & 3.0 & $2 \times 10^{4}$ & 0.56 & 1256.56  & 0.20 & 0.03\% & 2.64\%  \\

     &  & 0 & 10\%& 0 & 0.56 & $5\times 10^{4}$ & 16000  & 0.1 & 3.0 & $2 \times 10^{4}$ & 0.56 & 1256.56  & 0 & 0.01\% & 0\%  \\
     &  & 0 & 10\%& 0 & 0.56 & $5\times 10^{3}$ & 16000  & 0.1 & 3.0 & $2 \times 10^{4}$ & 0.56 & 1256.56  & 0.68 & 230.67\% & 100\%  \\ 

     &  & 0 & 5\%& 0 & 0.56 & $5\times 10^{4}$ & 16000  & 0.1 & 3.0 & $2 \times 10^{4}$ & 0.56 & 1256.56  & 0.05 & 0.01\% & 0.10\%  \\

     &  & 0 & 4\%& 0 & 0.56 & $5\times 10^{4}$ & 16000  & 0.1 & 3.0 & $2 \times 10^{4}$ & 0.56 & 1256.56  & 0.35 & 0.04\% & 10.89\%  \\ \\

 \hline 
	   \end{tabular}
    \begin{center}
  [Table continues on the next page]      
    \end{center}
  
    \begin{tablenotes}
    
        \item[(a),(b)] Initial spin period and initial angular frequency of rotation of the PNS respectively. For $B_0=0$\,G, the spin period is constant. 
        \item[(c)] Polar magnetic field strength of the PNS.
        \item[(d)] Periodic signal content.
        \item[(e)] Initial phase of the neutrino signal.
        \item[(f)] Neutrino luminosity decay coefficient.
        \item[(g),(h)] Total number of neutrino events and number of bins respectively.
        \item[(i),(j)] Time of arrival of first and last neutrino respectively.  
        \item[(k)] Number of PDF samples used to generate the neutrino data.
        \item[(l)] Detected neutrino luminosity decay coefficient.
        \item[(m)] Median of correctly detected angular frequencies.
        \item[(n)] Fraction of incorrect detections.
        \item[(o)] Median error.
        \item[(p)] Median false alarm rate.
        
    \end{tablenotes}
      \end{threeparttable}
\end{table*}

\begin{table*}
	\caption*{(Continuation of Table \ref{table1})}

	\begin{tabular}[width=\textwidth]{@{}cccccccccccccccc@{}} 
 
		\hline
		 $P_{\star0}  ^{\rm (a)}$ & $\Omega_{\star0} ^{(\rm b)}$ & $B_0  ^{(\rm c)}$ & M$ ^{(\rm d)}$ & $\phi_{i} ^{(\rm e)}$ & $\alpha ^{(\rm f)}$ & $\rm N_{\nu} ^{(\rm g)}$ & $\rm N_{\rm bins} ^{(\rm h)}$  & $t_i ^{(\rm i)}$ & $t_j ^{(\rm j)}$ & $\rm N_{\rm pdf} ^{(\rm k)}$  & $\alpha_{\rm b} ^{(\rm l)}$ & $\omega_{\rm p} ^{(\rm m)}$ & $F$ $ ^{(\rm n)}$ & $E$ $ ^{(\rm o)}$ &FAR$ ^{(\rm p)}$ \\ 
		(ms) &  (rad s$^{-1}$) & (G) & & & & & & (s) & (s) & & &   (rad s$^{-1}$) & & &
  \\
		\hline

\\

     5 & 1256.64 & 0 & 20\%& 0 & 0.56 & $5\times 10^{3}$ & 16000  & 0.1 & 3.0 & $4 \times 10^{4}$ & 0.56 & 2513.12  & 0 & 0.01\% & 0\%  \\

    &  & 0 & 15\%& 0 & 0.56 & $5\times 10^{4}$ & 16000  & 0.1 & 3.0 & $4 \times 10^{4}$ & 0.56 & 2513.12  & 0 & 0.01\% & 0\%  \\
    &  & 0 & 15\%& 0 & 0.56 & $5\times 10^{3}$ & 16000  & 0.1 & 3.0 & $4 \times 10^{4}$ & 0.56 & 2513.12  & 0 & 0.01\% & 0\%  \\

    &  & 0 & 13.5\%& 0 & 0.56 & $5\times 10^{3}$ & 16000  & 0.1 & 3.0 & $4 \times 10^{4}$ & 0.56 & 2513.51  & 0.20 & 0.01\% & 1.32\%  \\

    &  & 0 & 10\%& 0 & 0.56 & $5\times 10^{4}$ & 16000  & 0.1 & 3.0 & $4 \times 10^{4}$ & 0.56 & 2513.12  & 0 & 0.01\% & 0\%  \\
    &  & 0 & 10\%& 0 & 0.56 & $5\times 10^{3}$ & 16000  & 0.1 & 3.0 & $4 \times 10^{4}$ & 0.56 & 2513.12  & 0.64 & 64.64\% & 100\%  \\

    &  & 0 & 5\%& 0 & 0.56 & $5\times 10^{4}$ & 16000  & 0.1 & 3.0 & $4 \times 10^{4}$ & 0.56 & 2513.12  & 0.06 & 0.01\% & 0.10\%  \\

    &  & 0 & 4\%& 0 & 0.56 & $5\times 10^{4}$ & 16000  & 0.1 & 3.0 & $4 \times 10^{4}$ & 0.56 & 2513.12  & 0.26 & 0.01\% & 4.88\%  \\
    &  & 0 & 4\%& 0 & 0.56 & $5\times 10^{4}$ & 16000  & 0.1 & 3.0 & $8 \times 10^{4}$ & 0.56 & 2513.12  & 0.25 & 0.01\% & 4.39\%  \\
    &  & 0 & 4\%& 0 & 0.56 & $5\times 10^{4}$ & 32000  & 0.1 & 3.0 & $4 \times 10^{4}$ & 0.56 & 2513.51  & 0.27 & 0.01\% & 4.83\%  \\ \\

   1 & 6283.18 & 0 & 20\%& 0 & 0.56 & $5\times 10^{3}$ & 64000  & 0.1 & 3.0 & $2 \times 10^{5}$ & 0.56 & 12566.37  & 0 & 0\% & 0\%  \\

     &  & 0 & 15\%& 0 & 0.56 & $5\times 10^{4}$ & 64000  & 0.1 & 3.0 & $2 \times 10^{5}$ & 0.56 & 12566.37  & 0 & 0\% & 0\%  \\
     &  & 0 & 15\%& 0 & 0.56 & $5\times 10^{3}$ & 64000  & 0.1 & 3.0 & $2 \times 10^{5}$ & 0.56 & 12566.37  & 0.09 & 0\% & 0\%  \\

     &  & 0 & 13.5\%& 0 & 0.56 & $5\times 10^{3}$ & 64000  & 0.1 & 3.0 & $2 \times 10^{5}$ & 0.56 & 12566.37  & 0.21 & 0\% & 1.22\%  \\

     &  & 0 & 10\%& 0 & 0.56 & $5\times 10^{4}$ & 64000  & 0.1 & 3.0 & $2 \times 10^{5}$ & 0.56 & 12566.37  & 0 & 0\% & 0\%  \\
     &  & 0 & 10\%& 0 & 0.56 & $5\times 10^{3}$ & 64000  & 0.1 & 3.0 & $2 \times 10^{5}$ & 0.56 & 12566.37  & 0.59 & 14.89\% & 100\%  \\

     &  & 0 & 5\%& 0 & 0.56 & $5\times 10^{4}$ & 64000  & 0.1 & 3.0 & $2 \times 10^{5}$ & 0.56 & 12566.37  & 0.04 & 0\% & 0.10\%  \\
     &  & 0 & 5\%& 0 & 0.56 & $5\times 10^{4}$ & 64000  & 0.1 & 3.0 & $3 \times 10^{5}$ & 0.56 & 12566.37  & 0.04 & 0\% & 0.10\%  \\
     &  & 0 & 5\%& 0 & 0.56 & $5\times 10^{4}$ & 128000  & 0.1 & 3.0 & $2 \times 10^{5}$ & 0.56 & 12566.37  & 0.03 & 0\% & 0.10\%  \\
     &  & 0 & 4\%& 0 & 0.56 & $5\times 10^{4}$ & 64000  & 0.1 & 3.0 & $2 \times 10^{5}$ & 0.56 & 12566.37  & 0.29 & 0\% & 8.94\%  \\

\hline 
	   \end{tabular}
 \end{table*}

\begin{table*}
	\caption{Results from 3-parameter DFT. Refer to the footnotes in this Table and footnotes in Table \ref{table1} for the meaning of the quantities listed. All the results in this Table use $\rm N_{pdf}=2\times 10^{4}$. All the simulations except the one marked with \$ have $t_i=0.1$\,s and $t_f=3.0$\,s. All the results, except those marked and described in the table footnotes, are with a $\sin^2$ modulation as shown in Figure \ref{pdf_t3s} and left panel of Figure \ref{pdf_t3s_B2e15}. Simulations run with a $\sin^2$ modulation result in the largest peak in the Fourier power spectrum at twice the input signal frequency. Refer to Section \ref{alt_mod} and the caption of Figure \ref{pdf_t3s_B2e15} for the meaning of spindown rates labelled in terms of $B_0$ and $P_{\star0}$.}
	\label{table2}
      \begin{threeparttable}
      
	\begin{tabular}[width=\textwidth]{@{}ccccccccccccccc@{}} 
		\hline
		 $P_{\star0}$ & $[{\Omega_{\star i}},{\Omega_{\star f}}]^{(\rm a)}$ & $B_0$ & M & $\phi_{i}$ & $\alpha$ & $\rm N_{\nu}$ & $\rm N_{\rm bins}$     & $\alpha_{\rm b}$ & $\omega_{1\rm p}^{\rm (b)}$ & $\omega_{2\rm p}^{\rm (c)}$ & $f_{3\rm p}^{\rm (d)}$ & $F$ & $E$ &FAR \\ 
		(ms) & (rad s$^{-1}$) & (G) & & & & &   &  & (rad s$^{-1}$) & (rad s$^{-1}$) & (s$^{-1}$)\\
		\hline

\\
200  & [31.42, 31.42] & 0 & 15\%& 0 & 0.56 & $5\times 10^{3}$ & 800  & 0.56 & 6.28 & 56.44 & -0.001 & 0.09 & 0.78\% & 0\%  \\

   &  & 0 & 14\%& 0 & 0.56 & $5\times 10^{3}$ & 800   & 0.56 & 6.28 & 56.44 & -0.001 & 0.18 & 0.97\% & 0.24\%  \\

   &  & 0 & 14\%& $\pi$/6 & 0.56 & $5\times 10^{3}$ & 800  &  0.57 & 6.28 & 56.44 & -0.001 & 0.21 & 1.02\% & 0.68\%  \\
   &  & 0 & 14\%& $\pi$/4 & 0.56 & $5\times 10^{3}$ & 800  & 0.57 & 6.28 & 56.44 & -0.001 & 0.22 & 1.05\% & 1.32\%  \\
   &  & 0 & 14\%& $\pi$/3 & 0.56 & $5\times 10^{3}$ & 800  & 0.57 & 6.28 & 56.44 & -0.001 & 0.21 & 0.96\% & 0.98\%  \\
   &  & 0 & 14\%& $\pi$/2 & 0.56 & $5\times 10^{3}$ & 800 & 0.56 & 6.28 & 56.44 & -0.001 & 0.15 & 0.92\% & 0.19\%  \\
   &  & 0 & 14\%& 0 & 0.56 & $5\times 10^{3}$ & 1600  & 0.56 & 6.28 & 56.44 & -0.001 & 0.18 & 0.94\% & 0.29\%  \\
   &  & $^{\sim}$0 & 14\%& $5\pi/6$ & 0.56 & $5\times 10^{3}$ & 800  & 0.54 & 22.49 & 8.60 & -0.017 & 0 & 0.98\% & 0\%  \\

   &  & 0 & 13.5\%& 0 & 0.56 & $5\times 10^{3}$ & 800  & 0.56 & 6.28 & 56.44 & -0.001 & 0.25 & 1.15\% & 1.66\%  \\

   &  & 0 & 10\%& 0 & 0.56 & $5\times 10^{4}$ & 800  & 0.56 & 6.28 & 56.44 & -0.001 & 0 & 0.40\% & 0\%  \\
   &  & 0 & 10\%& 0 & 0.56 & $5\times 10^{3}$ & 800  & 0.56 & 6.28 & 56.44 & -0.001 & 0.68 & 112.79\% & 100\%  \\

   &  & 0 & 5\%& 0 & 0.56 & $5\times 10^{4}$ & 800  & 0.56 & 6.28 & 56.44 & -0.001 & 0.05 & 0.75\% & 0.19\%  \\

   &  & 0 & 4\%& 0 & 0.56 & $5\times 10^{4}$ & 800  & 0.56 & 6.28 & 56.44 & -0.001 & 0.35 & 1.45\% & 10.94\%  \\ \\

50 & [125.66, 77.26] & $10^{15}$ & 15\%& 0 & 0.56 & $5\times 10^{3}$ & 800 & 0.56 & 83.45 & 174.50 & 0.313 & 0.18 & 0.94\% & 0.19\%  \\

 &  & $10^{15}$ & 13.5\%& 0 & 0.56 & $5\times 10^{3}$ & 800 & 0.56 & 78.82 & 177.58 & 0.297 & 0.34 & 1.16\% & 3.71\%  \\

 &  & $10^{15}$ & 10\%& 0 & 0.56 & $5\times 10^{4}$ & 800 & 0.56 & 93.48 & 166.78 & 0.346 & 0 & 0.71\% & 0\%  \\
 &  & $10^{15}$ & 10\%& 0 & 0.56 & $5\times 10^{3}$ & 800  & 0.56 & 88.07 & 171.41 & 0.330 & 0.75 & 20.50\% & 100\%  \\
  &  & $10^{15}$ & 5\%& 0 & 0.56 & $5\times 10^{4}$ & 800   & 0.56 & 94.25 & 165.23 & 0.346 & 0.10 & 0.84\% & 0\%  \\
  &  & $10^{15}$ & 4\%& 0 & 0.56 & $5\times 10^{4}$ & 800  & 0.56 & 93.47 & 166.78 & 0.346 & 0.40 & 1.49\% & 13.77\%  \\
  &  & $10^{15}$ & 4\%& 0 & 0.56 & $5\times 10^{4}$ & 1600  & 0.56 & 71.87 & 184.53 & 0.280 & 0.38 & 1.38\% & 6.35\%  \\
  &  & $10^{15}$ & 4\%& 0 & 0.56 & $5\times 10^{4}$ & 3200  & 0.56 & 93.48 & 166.78 & 0.346 & 0.38 & 1.52\% & 10.30\%  \\

\\
200 & [31.42, 12.57] & $2\times10^{15}$ & 15\%& 0 & 0.56 & $5\times 10^{3}$ & 800 & 0.56 & 18.63 & 44.09 & 0.644 & 0.18 & 2.47\% & 0.10\%  \\
 &  & $2\times10^{15}$ & 14\%& 0 & 0.56 & $5\times 10^{4}$ & 800  & 0.56 & 18.63 & 44.09 & 0.644 & 0 & 1.09\% & 0\%  \\
 &  & $2\times10^{15}$ & 14\%& 0 & 0.56 & $5\times 10^{3}$ & 800  & 0.56 & 17.86 & 44.86 & 0.628 & 0.26 & 2.79\% & 0.78\%  \\

&  & $2\times10^{15}$ & 13.5\%& 0 & 0.56 & $5\times 10^{3}$ & 800  & 0.56 & 18.63 & 43.32 & 0.644 & 0.29 & 2.92\% & 2.83\%  \\

 &  & $2\times10^{15}$ & 10\%& 0 & 0.56 & $5\times 10^{4}$ & 800  & 0.56 & 18.63 & 44.09 & 0.644 & 0 & 1.23\% & 0\%  \\
 &  & $2\times10^{15}$ & 10\%& 0 & 0.56 & $5\times 10^{3}$ & 800  & 0.56 & 17.86 & 44.86 & 0.628 & 0.75 & 293.67\% & 100\%  \\
 &  & $2\times10^{15}$ & 5\%& 0 & 0.56 & $5\times 10^{4}$ & 800  & 0.56 & 18.63 & 44.09 & 0.644 & 0.13 & 2.31\% & 0.10\%  \\
 &  & $2\times10^{15}$ & 4\%& 0 & 0.56 & $5\times 10^{4}$ & 800 & 0.56 & 20.17 & 42.55 & 0.694 & 0.40 & 3.58\% & 13.04\%  \\
&  & $2\times10^{15}$ & 4\%& 0 & 0.56 & $5\times 10^{4}$ & 1600 & 0.56 & 20.17 & 42.55 & 0.694 & 0.39 & 3.39\% & 12.11\%  \\
 &  & $^{@}2\times10^{15}$ & 4\%& 0 & 0.56 & $5\times 10^{4}$ & 800 & 0.56 & 18.61 & 44.05 & 0.645 & 0.40 & 3.41\% & 12.30\%  \\

  &  & $2\times10^{15}$ & 10\%& 0 & 0.8 & $5\times 10^{4}$ & 800  & 0.79 & 18.63 & 44.09 & 0.644 & 0 & 1.23\% & 0\%  \\
  &  & $2\times10^{15}$ & 5\%& 0 & 0.8 & $5\times 10^{4}$ & 800 & 0.80 & 20.17 & 42.55 & 0.694 & 0.16 & 2.47\% & 0.24\%  \\
 &  & $2\times10^{15}$ & 10\%& 0 & $h(t)^{*}$ & $5\times 10^{4}$ & 800 & $g(t)^{*}$ & 18.63 & 44.09 & 0.644 & 0 & 1.32\% & 0\%  \\
 &  & $2\times10^{15}$ & 5\%& 0 & $h(t)^{*}$ & $5\times 10^{4}$ & 800  & $g(t)^{*}$ & 17.86 & 44.09 & 0.611 & 0.16 & 2.47\% & 0\%  \\
 &  & $^{\#}2\times10^{15}$ & 10\%& 0 & 0.56 & $5\times 10^{4}$ & 800  & 0.59 & 18.63 & 44.09 & 0.644 & 0.20 & 0.84\% & 0\%  \\
 &  & $^{\#}2\times10^{15}$ & 5\%& 0 & 0.56 & $5\times 10^{4}$ & 800  & 0.58 & 18.63 & 44.09 & 0.644 & 0.38 & 1.62\% & 0\%  \\
 &  & $^{\&}2\times10^{15}$ & 5\%& 0 & 0.56 & $5\times 10^{4}$ & 800 & 0.57 & 9.37 & 21.72 & 0.644 & 0.12 & 2.58\% & 0\%  \\
 &  & $^{\&\$}2\times10^{15}$ & 5\%& 0 & 0.56 & $5\times 10^{4}$ & 800 & 0.57 & 9.37 & 21.72 & 0.644 & 0.09 & 2.18\% & 0\%  \\ 

  &  & $^{\sim}2\times10^{15}$ & 5\%& $5\pi/6$ & 0.56 & $5\times 10^{4}$ & 800 & 0.55 & 10.14 & 21.72 & 0.710 & 0.27 & 4.31\% & 0\%  \\ \\

 \hline 
	   \end{tabular}
        \begin{center}
  [Table continues on the next page]      
    \end{center}
  \begin{tablenotes}
        \item[$\sim$] Simulation with a physically motivated modulation function as shown in Figure \ref{pdf_t3s_Chrismod}. For this modulating function, the input signal frequency is equal to the PNS frequency of rotation.
        \item[@] Simulation with number of grid points $(N_{\omega_1},N_{\omega_2},N_{f_3})=(800,800,32)$ while all others are with $(N_{\omega_1},N_{\omega_2},N_{f_3})=(400,400,128)$.
        \item[*]Simulation  with a time-dependent neutrino luminosity decay coefficient. We set $h(t)=0.56$ for $t_i\le t\le0.5$\,s and $h(t)=0.56+0.10(t-0.5\,s)$ for $0.5\,{\rm s}<t\le t_f$. From the power law fit and averaging over 250 sets of neutrino counts (see Sections \ref{search_algos} and \ref{far_sub} for details), we find $g(t)=0.55$ with $\rm M=10\%$ and $g(t)=0.56$ with $\rm M=5\%$ for $t_i\le t\le0.5$\,s and $g(t)=0.10(t-0.5$\,s) for $0.5\,{\rm s}<t\le t_f$. 
        \item[\#] Simulation with a square wave modulation with a duty cycle of 10\% as shown in the middle panel of Figure \ref{pdf_t3s_B2e15}.
        \item[\&] Simulation with a modulating function obtained from combination of sinusoidal and square waves as shown in the right panel Figure \ref{pdf_t3s_B2e15}. For this modulating function, the input signal frequency is equal to the PNS frequency of rotation.  
        \item[\$] Simulation with $t_i=1.0$\,s and $t_f=3.9$\,s. 
        \item[(a)] $\Omega_{\star i}$ and $\Omega_{\star f}$ are the PNS angular frequencies at $t=t_i$ and $t=t_f$ respectively. The PNS angular frequency is constant as a function of time for $B_0=0$\,G.
        \item[(b),(c),(d)] Frequencies at the maximum of the Fourier power spectrum that result in minimum deviation from the input signal frequency out of the 250 different sets \\ of neutrino counts (see Section \ref{spindown_results} for details).
   \end{tablenotes}     
      \end{threeparttable}
\end{table*}

\begin{table*}
	\caption*{(Continuation of Table \ref{table2})}

	\begin{tabular}[width=\textwidth]{@{}ccccccccccccccc@{}} 
		\hline
		 $P_{\star0}$ & $[{\Omega_{\star i}},{\Omega_{\star f}}]^{(\rm a)}$ & $B_0$ & M & $\phi_{i}$ & $\alpha$ & $\rm N_{\nu}$ & $\rm N_{\rm bins}$     & $\alpha_{\rm b}$ & $\omega_{1\rm p}^{\rm (b)}$ & $\omega_{2\rm p}^{\rm (c)}$ & $f_{3\rm p}^{\rm (d)}$ & $F$ & $E$ &FAR \\ 
		(ms) & (rad s$^{-1}$) & (G) & & & & &   &  & (rad s$^{-1}$) & (rad s$^{-1}$) & (s$^{-1}$)\\
		\hline

\\

400 & [15.71, 3.07] & $4\times10^{15}$ & 20\%& 0 & 0.56 & $5\times 10^{3}$ & 800 & 0.55 & 6.28 & 24.80 & 1.157 & 0.28 & 4.20\% & 0\%  \\
 & & $4\times10^{15}$ & 15\%& 0 & 0.56 & $5\times 10^{3}$ & 800  & 0.55 & 6.28 & 24.80 & 1.157 & 0.45 & 4.83\% & 1.76\%  \\
 & & $4\times10^{15}$ & 10\%& 0 & 0.56 & $5\times 10^{4}$ & 800 & 0.56 & 6.28 & 24.80 & 1.157 & 0.01 & 3.69\% & 0\%  \\
 & & $4\times10^{15}$ & 10\%& 0 & 0.56 & $5\times 10^{3}$ & 800  & 0.56 & 6.28 & 24.03 & 1.124 & 0.83 & 34.67\% & 100\%  \\
& & $4\times10^{15}$ & 5\%& 0 & 0.56 & $5\times 10^{4}$ & 800  & 0.56 & 6.28 & 24.80 & 1.157 & 0.20 & 4.17\% & 0\%  \\
& & $4\times10^{15}$ & 4\%& 0 & 0.56 & $5\times 10^{4}$ & 800 & 0.56 & 6.28 & 24.80 & 1.157 & 0.33 & 4.47\% & 0.49\%  \\
& & $4\times10^{15}$ & 3\%& 0 & 0.56 & $5\times 10^{4}$ & 800  & 0.56 & 6.28 & 24.80 & 1.157 & 0.60 & 6.38\% & 100\%  \\

 \hline 
	   \end{tabular}
 
\end{table*}

 \begin{figure}
\centering{}
\includegraphics[width=\linewidth]{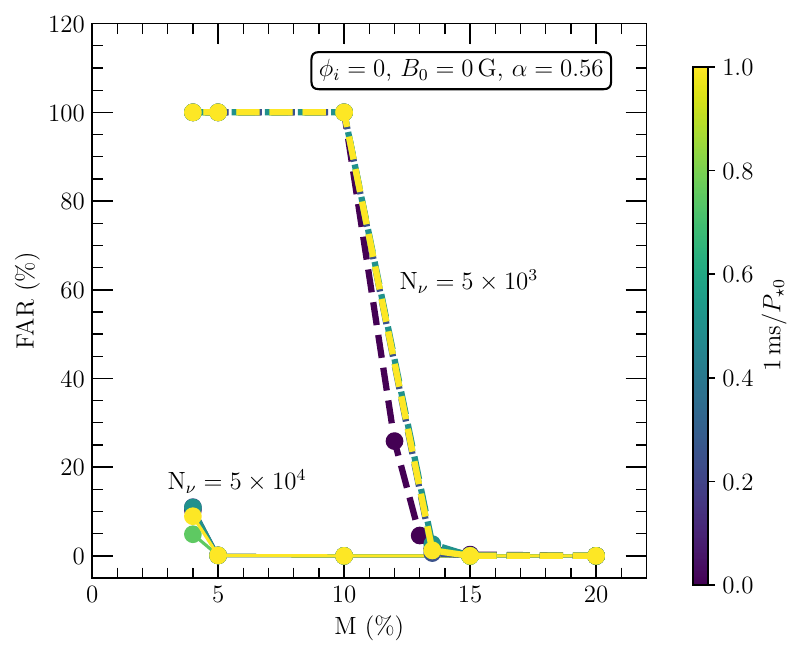}
\caption{False alarm rate (FAR) as a function of periodic signal content and PNS spin period $P_{\star0}$ (in ms). The colors are normalized to 1\,ms, meaning that yellow corresponds to $P_{\star0}=1$\,ms and dark violet corresponds to $P_{\star0}=200$\,ms. 
This figure shows results from Table \ref{table1}. Since the PNS spin period is constant as a function of time at polar magnetic field strength $B_0=0$\,G (refer to the caption of Figure \ref{pdf_t3s}), 1-parameter DFT is sufficient. All the results in this figure are with a $\sin^2$ modulation.} 
\label{FAR_1d_fig}
\end{figure}

\begin{figure*}
\centering{}
\includegraphics[width=\textwidth]{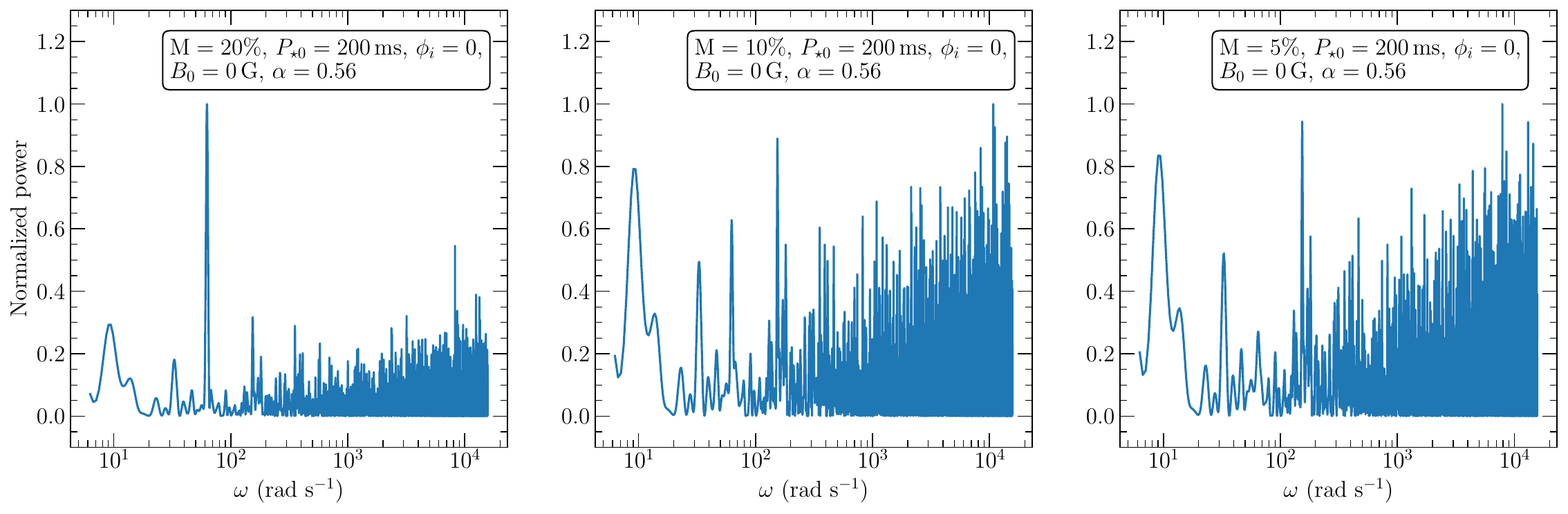}
\caption{Normalized one-parameter Fourier power spectrum using a $\sin^2$ modulation at a constant PNS spin period of 200\,ms with a total of $5\times 10^{3}$ neutrinos for various values of periodic content M. We use the label polar magnetic field $B_0=0$\,G to imply a constant PNS spin period (refer to the caption of Figure \ref{pdf_t3s}).} 
\label{1d_DFT_5e3}
\end{figure*}

 \begin{figure*}
\centering{}
\includegraphics[width=\textwidth]{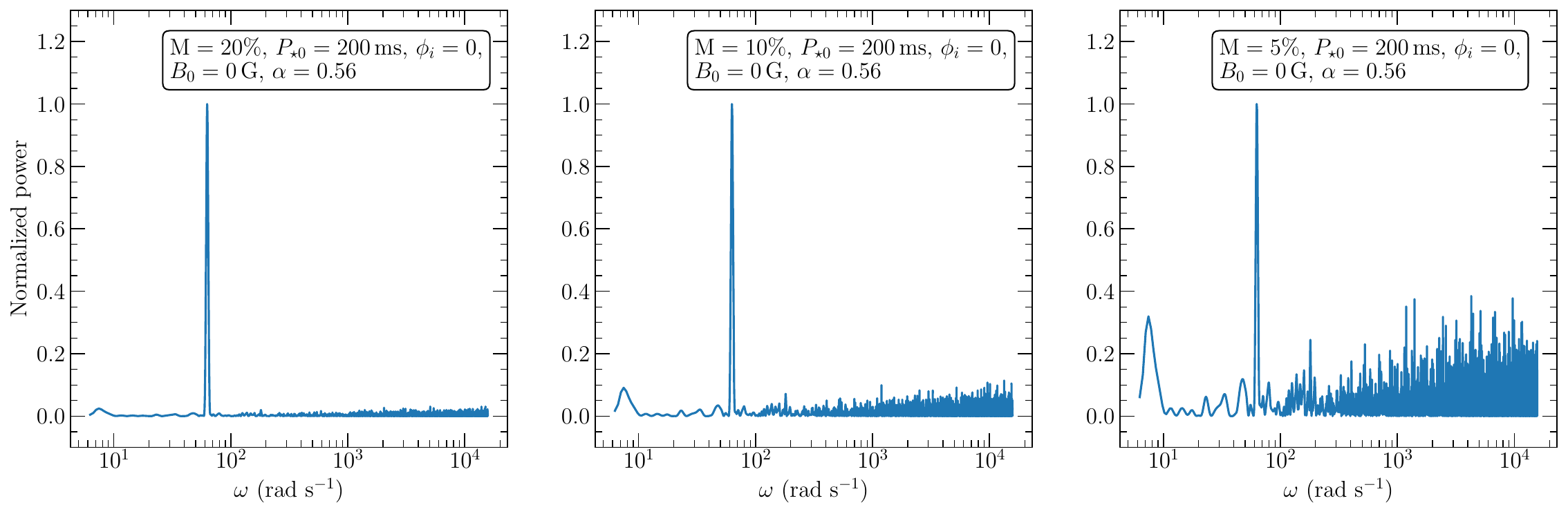}
\caption{Same as Figure \ref{1d_DFT_5e3}, but for a total of $5\times 10^{4}$ neutrinos.} 
\label{1d_DFT_5e4}
\end{figure*}

 \begin{figure*}
\centering{}
\includegraphics[width=\textwidth]{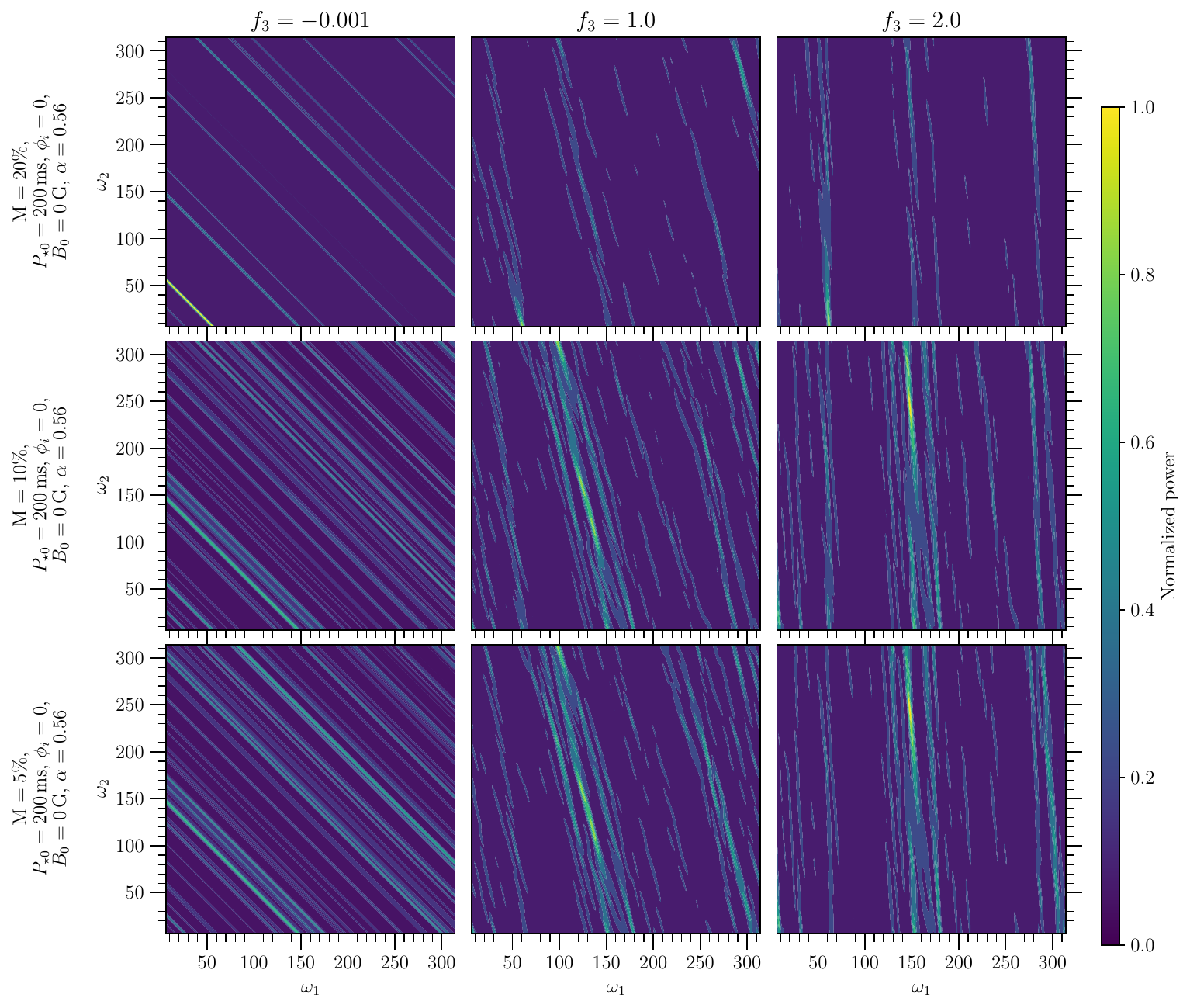}
\caption{Normalized 3-parameter Fourier power spectrum at a constant PNS spin period of 200\,ms with a total of $5\times 10^{3}$ neutrinos for various values of periodic content M for a $\sin^2$ modulation. The normalized power spectrum is shown as a function of $\omega_1$ and $\omega_2$ at three different values of $f_3$. We use the label polar magnetic field $B_0=0$\,G to imply a constant PNS spin period (refer to the caption of Figure \ref{pdf_t3s}).} 
\label{3d_DFT_B0_5e3}
\end{figure*}

 \begin{figure*}
\centering{}
\includegraphics[width=\textwidth]{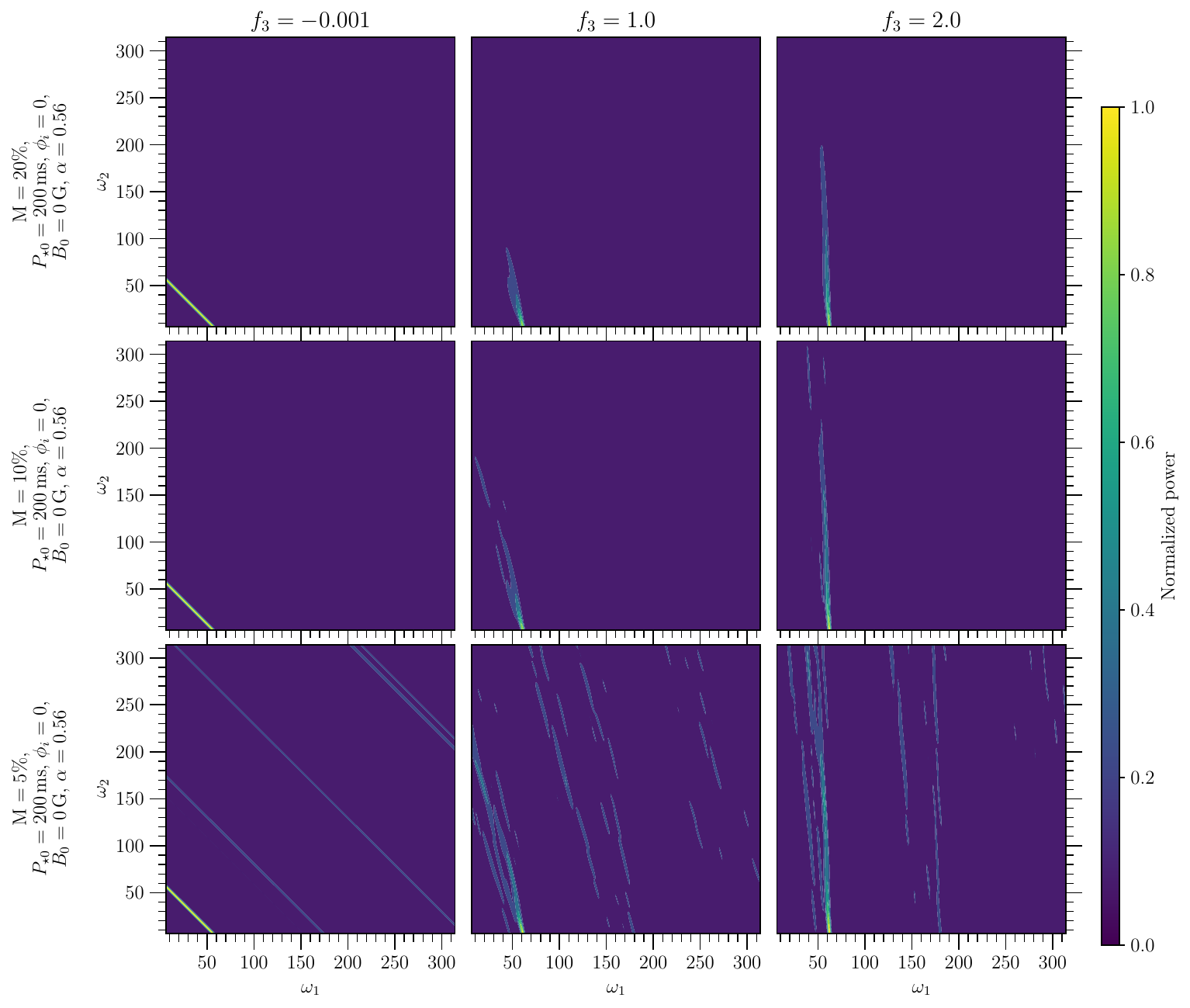}
\caption{Same as Figure \ref{3d_DFT_B0_5e3}, but with $5\times10^{4}$ neutrinos.} 
\label{3d_DFT_B0_5e4}
\end{figure*}

 \begin{figure*}
\centering{}
\includegraphics[width=\textwidth]{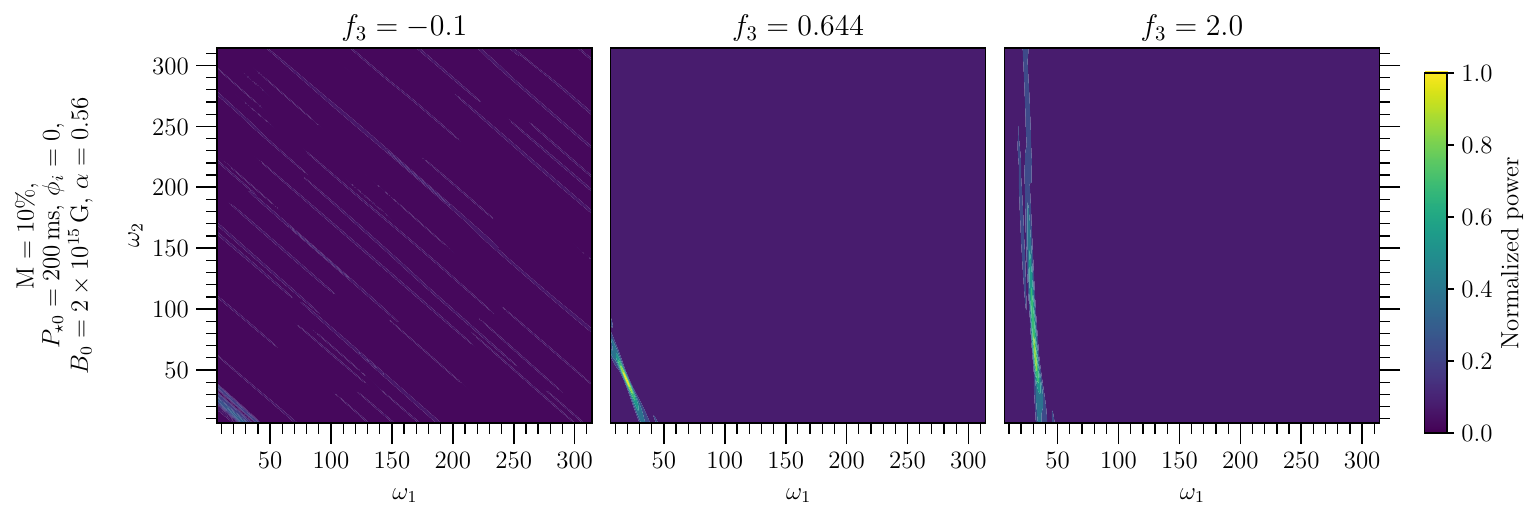}
\caption{Normalized 3-parameter Fourier power spectrum for a $\sin^2$ modulation at an initial PNS spin period $P_{\star0}=200$\,ms and polar magnetic field strength $B_0=2\times10^{15}$\,G with a total of $5\times 10^{4}$ neutrinos for periodic content $\rm M=10\%$. Refer to the caption of Figure \ref{pdf_t3s_B2e15} for the meaning of spindown rates labelled in terms of $B_0$ and $P_{\star0}$.} 
\label{3d_DFT_B2e15_5e4}
\end{figure*}

 \begin{figure*}
\centering{}
\includegraphics[width=\textwidth]{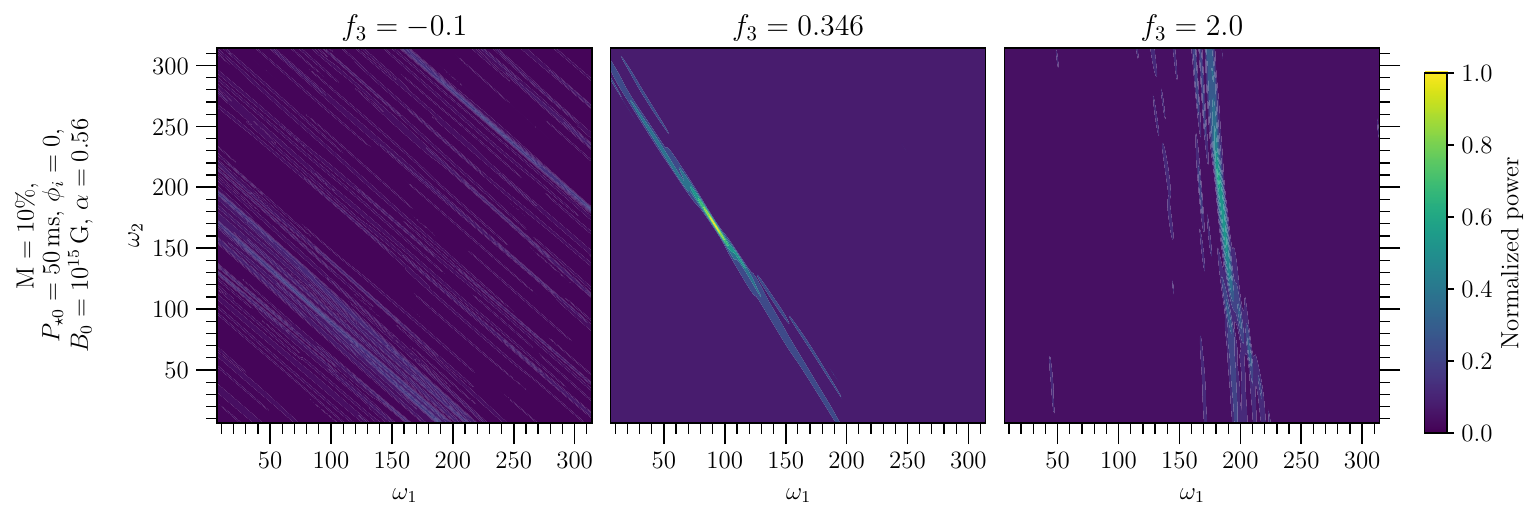}
\caption{Normalized 3-parameter Fourier power spectrum for a $\sin^2$ modulation at an initial PNS spin period $P_{\star0}=50$\,ms and polar magnetic field strength $B_0=10^{15}$\,G with a total of $5\times 10^{4}$ neutrinos for periodic content $\rm M=10\%$. Refer to the caption of Figure \ref{pdf_t3s_B2e15} for the meaning of spindown rates labelled in terms of $B_0$ and $P_{\star0}$.} 
\label{3d_DFT_B1e15_5e4}
\end{figure*}

 \begin{figure*}
\centering{}
\includegraphics[width=\textwidth]{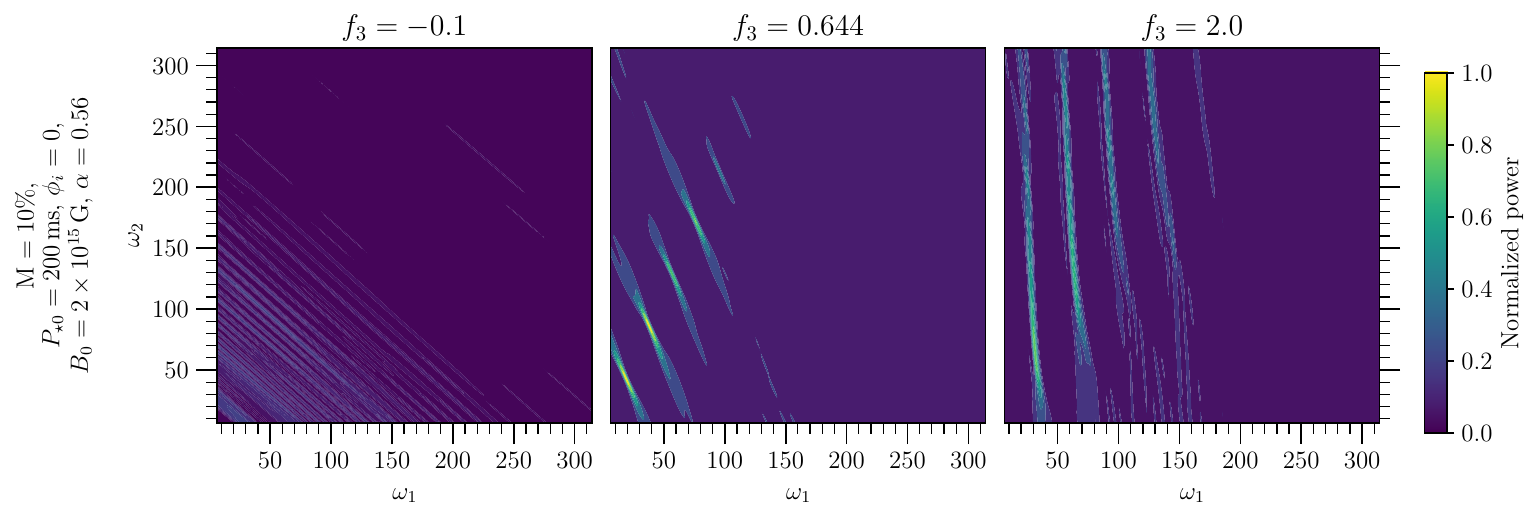}
\caption{Same as Figure \ref{3d_DFT_B2e15_5e4}, but with a square wave modulation as shown in the middle panel of Figure \ref{pdf_t3s_B2e15}. Although the largest peak occurs at the fundamental frequency parameters of the input signal, there are peaks at higher harmonics as well.} 
\label{3d_DFT_B2e15_5e4_sqwave}
\end{figure*}

Figure \ref{3d_DFT_B0_5e3} shows the normalized 3-parameter Fourier power spectrum (see equations \ref{3d_DFT} and \ref{3d_power}) at a constant PNS spin period of 200\,ms for various values of periodic content M with $5\times 10^{3}$ neutrinos. At $\rm M=20\%$, the maximum of the power spectrum occurs at $f_3\approx 0$ and along the line $\omega_1+\omega_2=62.81$\,rad s$^{-1}$ because there is no spindown. The detected frequency is twice the PNS frequency of rotation because we use a $\sin^2$ modulation. As the value of M decreases, stronger peaks occur at frequencies different from the input signal frequency. Figure \ref{3d_DFT_B0_5e4} is with the same parameters as Figure \ref{3d_DFT_B0_5e3}, but with $5\times 10^4$ neutrinos. As the number of neutrinos increase, the input signal strength increases and the maximum of the power spectrum occurs at the input signal frequencies for lower values of M as well.  Figure \ref{3d_DFT_B2e15_5e4} shows the 3-parameter power spectrum for a PNS with an initial spin period $P_{\star0}=200$\,ms and polar magnetic field strength $B_0=2\times 10^{15}$\,G with a total of $5\times10^{4}$ neutrinos. Since the PNS spins down with time in this case, the maximum of the power spectrum occurs at $f_3>0$ and is localised on the $\omega_1$ and $\omega_2$ plane. Figure \ref{3d_DFT_B1e15_5e4} shows the power spectrum with $5\times 10^{4}$ neutrinos for a PNS with $P_{\star0}=50$\,ms and $B_0=10^{15}$\,G. As expected, we find that the spindown parameter $f_3$ increases as $B_0$ and $P_{\star0}$ increase (see also Table \ref{table2}). A $\sin^2$ modulation is used for the simulations in Figures \ref{3d_DFT_B2e15_5e4} and \ref{3d_DFT_B1e15_5e4}. Figure \ref{3d_DFT_B2e15_5e4_sqwave} is the same as Figure \ref{3d_DFT_B2e15_5e4}, but with a square wave modulation as shown in the middle panel of Figure \ref{pdf_t3s_B2e15}. Since the modulating function is a square wave, there are peaks in the Fourier power spectrum at higher harmonics as well. The largest peak still occurs at the fundamental frequency of the input signal.

\subsection{Effects of de-trending the neutrino signal}
\label{detrend}
As shown in Figure \ref{pons_lum}, the neutrino luminosity decays with time which results in a decaying amplitude in the PDF (see Figures \ref{pdf_t3s}, \ref{pdf_t3s_B2e15} and \ref{pdf_t3s_Chrismod}) and the neutrino counts (see Figures \ref{counts_5e3_t3s} and \ref{counts_5e4_t3s}). This decaying trend can give rise to spurious peaks in the Fourier power spectrum. Multiplying the neutrino counts with a window function (see Section \ref{bins}) removes the jump discontinuity at the edges and minimizes the magnitude of the spurious peaks. In principle, the neutrino counts data can be processed further before applying the window function to further minimize the peaks due to the trend in the signal. We can fit a function to the neutrino counts data and subtract it out to remove the trend. If the function well approximates the trend in the input signal, then this can lead to reduction in the false alarm rates. 

To test this, we have run simulations by fitting and subtracting out polynomials of order 2, 3, 4 and 5 to remove the trend in the neutrino counts data before applying the window. Since the neutrino luminosity decays as a power law, we have also run simulations by fitting and subtracting out a power law. We apply this de-trending procedure to the neutrino counts data with periodic signal as well as the pure power law signal used to estimate FAR. We test the effects of this procedure for the one-parameter DFT. From Table \ref{table1}, for a constant PNS spin period of $P_{\star0}=200$\,ms and periodic content $\rm M=4\%$, the false alarm rate (FAR) is $10.35\%$ with $5\times 10^{4}$ neutrinos. Subtracting polynomials of order 2, 3, 4, and 5 to remove the trend in the input signal results in FAR of 6.45\%, 6.15\%, 14.01\% and 20.90\% respectively. Fitting and subtracting out a power law results in FAR of 6.05\%. We find that subtracting out a power law, quadratic or cubic polynomials from the neutrino counts data marginally improves FAR, but not significantly enough to change the conclusion about a significant detection. With $5\times 10^{3}$ neutrinos and $\rm M=13\%$, we find that the FAR remains the same after the de-trending procedure with a power law and polynomials of order 2, 3, 4, and 5. Thus, we present results without the de-trending procedure in this paper. 

\section{Discussion \& Conclusions}
\label{conclusions}
In this paper, we explore the prospects for detecting proto-neutron star (PNS) rotation and spindown using supernova neutrinos as messengers. Provided there is long-lived neutrino emission asymmetry on the PNS lasting a few seconds during the cooling epoch, we can expect a periodic modulation in neutrino counts per time interval. As mentioned in Section \ref{section:introduction}, asymmetries such as LESA or magnetar strength magnetic fields can give rise to asymmetric neutrino emission. Irrespective of the physics of the asymmetry, provided that there are roughly stable neutrino emission hot-spots or cold-spots on the PNS, we explore the possibility of using the neutrino arrival times to detect PNS rotation rates. Due to lack of neutrino arrival times from a Galactic supernova (except $\sim 20$ from SN1987A, see \citealt{Hirata1987,Bionta1987}), we use toy models to generate neutrino arrival times based on a probability distribution (see equation \ref{PDF}). The toy models we use consist of a non-periodic part and a periodic modulation both of which decay as a power law with time to mimic the drop in neutrino luminosity as a function of time (see Figures \ref{pons_lum}, \ref{pdf_t3s} and \ref{pdf_t3s_B2e15}). The neutrino luminosity decay profiles we use to model the cooling phase (see Figure \ref{pons_lum}) are motivated by \cite{Pons1999} cooling models.

We use the transformation method \citep{Numerical_recipes} to generate neutrino arrival times according to a given PDF $P(t)$ (see Section \ref{arrival_times_subsub} for details). We distribute the neutrino arrival times into uniform bins to obtain the neutrino counts as a function of time. Provided there are hot-spots on the PNS, the neutrino counts exhibit periodic modulation as shown in Figures \ref{counts_5e3_t3s} and \ref{counts_5e4_t3s}. If the PNS spin period is constant as a function of time, then a regular 1-parameter Discrete Fourier Transform (DFT) is sufficient to extract the frequency content from the neutrino signal. If the PNS happens to be a magnetar with polar magnetic field strength $B_0\gtrsim 10^{15}$\,G and is born `slowly' rotating with initial spin period $P_{\star0}\gtrsim 100$\,ms, then \cite{Prasanna2022} show that the PNS can spindown to spin periods of the order of seconds during the PNS cooling phase lasting just a few tens of seconds. Motivated by such a scenario, we propose a modified 3-parameter DFT technique to detect PNS spindown (see Section \ref{search_algos}). We characterize the significance of the peaks in the Fourier power spectrum using the false alarm rate (FAR) as described in Section \ref{far_sub}.

Table \ref{table1} shows the results from the 1-parameter Fourier power spectrum (see equations \ref{1d_DFT} and \ref{1d_power}). Figure \ref{FAR_1d_fig} shows the variation of FAR obtained using the 1-parameter DFT as a function of the periodic content M (equation \ref{per_cont}) for various spin periods. We show that all PNS spin periods can be detected with $\rm FAR<1\%$ with a total of $5\times 10^{4}$ neutrinos in $\sim 3$\,s for periodic signal content $\rm M\geq5\%$. With a total of $5\times10^{3}$ neutrinos in $\sim 3$\,s, detection is possible with $\rm FAR <2\%$ for periodic signal content $\gtrsim13\%$. Figures \ref{1d_DFT_5e3} and \ref{1d_DFT_5e4} show examples of 1-parameter Fourier power spectrum with $5\times10^{3}$ and $5\times10^{4}$ neutrinos respectively.

Table \ref{table2} shows results from the 3-parameter Fourier power spectrum (see equations \ref{3d_DFT} and \ref{3d_power}). Figure \ref{FAR_2d_fig} shows the variation of FAR obtained using the 3-parameter DFT as a function of the periodic content M for various spindown rates. We show that PNS spindown can be detected with $\rm FAR<1\%$ for $\rm M\geq 5\%$ with $5\times 10^{4}$ neutrinos and with $\rm FAR<2\%$ for $\rm M\gtrsim 13-15\%$ with $5\times 10^{3}$ neutrinos in $\sim 3$\,s. We show that spindown detection is possible with various neutrino luminosity decay profiles. We also show that detection is not limited to a specific choice of the modulating function by testing our detection algorithm for various modulating functions listed in Table \ref{mod_func_table}. Figures \ref{3d_DFT_B0_5e3}-\ref{3d_DFT_B2e15_5e4_sqwave} show the normalized 3-parameter Fourier power spectrum for various values of initial spin period $P_{\star0}$, polar magnetic field strength $B_0$, periodic content M, and modulating functions. 

The neutrino detection efficiency in the neutrino detectors can decrease as the neutrino energies decrease with time (e.g., \citealt{Pons1999, Li2021}). This might result in a steeper power law at late times. While most of the results in this paper use neutrino luminosity decay coefficient $\alpha=0.56$, we have presented some results with $\alpha=0.8$ and also a time dependent $\alpha$ to show that detection is possible in these cases as well (see Table \ref{table2}). With a much steeper power law decay in neutrino luminosity, larger number of neutrinos may be required for detection of PNS rotation rate and spindown for a given value of periodic signal content M.

For the modulating functions used in this paper (see Table \ref{mod_func_table}), the largest peak in the Fourier power spectrum occurs at the fundamental frequency of the input signal. It is possible to construct modulating functions such that the largest peak in the power spectrum occurs at some multiple of the fundamental signal frequency. If there are multiple peaks with a similar magnitude in the power spectrum (like in the case of square wave modulation considered in this paper, see Figure \ref{3d_DFT_B2e15_5e4_sqwave}), it may be difficult to get the fundamental frequency of the input signal and hence the PNS rotation rates without detailed modeling of the neutrinos emission asymmetries. In such a scenario, we can only identify some multiple of the PNS rotation rate from the power spectrum. However, for the spindown model we use (equation \ref{omega_model}), the spindown parameter $f_3$ remains the same for all the peaks corresponding to the various harmonics. Hence, in the cases of spindown with $f_3>0$, we may be able to use spindown models from \cite{Prasanna2022,Prasanna2023} to match the likely initial spin period of the PNS with the spindown parameter. 

We note that the techniques described in this paper cannot be applied to the neutrino data from SN1987A since only $\sim 20$ neutrinos were detected. The signal strength is too low with such a small number of neutrinos. However, attempts have been made to extract PNS spin period from the neutrino arrival times even with such a small number of neutrinos (e.g., \citealt{Harwit1987}), although these methods have been contested (e.g., \citealt{Schaefer1988}). With modern detectors like Super-Kamiokande, about $10^{4}$ neutrino detections can be expected from a supernova at a distance of 10\,kpc \citep{Beacom2000, Beacom2002}. The number of neutrino detections increase to $\sim 5-9\times 10^{4}$ in a bigger detector like Hyper-Kamiokande \citep{Abe2021} for a supernova at 10\,kpc. When a Galactic supernova occurs, if we get a shower of a large number neutrinos in $\sim 1-100$\,s of the PNS cooling phase, our results show that we can detect PNS rotation and spindown parameters if there are stable neutrino emission hot-spots or cold-spots on the PNS resulting in a $\sim 5-15\%$ periodic modulation in neutrino counts.

\section*{Acknowledgments}
\label{section:acknowledgements}
We thank John Beacom and Christopher Kochanek for helpful discussions. TP and TAT are supported in part by NASA grant 80NSSC20K0531. CH was supported by NASA award 15-WFIRST15-0008, Simons Foundation award 60052667, and the David \& Lucile Packard Foundation. We have run our simulations on the Ohio supercomputer \citep{OhioSupercomputerCenter1987}.

\section*{Data Availability}
The code to reproduce the results in this paper is available upon request. 


\bibliographystyle{mnras}
\bibliography{ref} 

\bsp	
\label{lastpage}
\end{document}